\documentclass[conference]{IEEEtran}
\IEEEoverridecommandlockouts 
\usepackage{cite}
\usepackage{amsmath,amssymb}
\usepackage{graphicx}
\usepackage{booktabs,multirow,array}
\usepackage[ruled,vlined,linesnumbered]{algorithm2e}
\SetAlFnt{\small}
\SetAlCapFnt{\small}
\SetAlCapNameFnt{\small}
\SetAlgoNlRelativeSize{-1}
\SetAlgoHangIndent{1em}
\DontPrintSemicolon
\usepackage[table]{xcolor}
\usepackage[hidelinks]{hyperref}

\definecolor{resulthead}{HTML}{ECEEF1}
\definecolor{resultsubhead}{HTML}{F5F6F8}
\definecolor{bestfill}{HTML}{B5D2EE}
\definecolor{secondfill}{HTML}{E4EEF9}
\definecolor{resultdivider}{HTML}{D4D9DE}

\newtheorem{definition}{Definition}

\newcommand{\method}{GenTUS}

\title{GenTUS: Table Union Search via Constrained Generation of Semantic Table Identifiers}
\author{
  \IEEEauthorblockN{Shulun Zhang\textsuperscript{1},
    Linting Wang\textsuperscript{2},
    Yuwei Xu\textsuperscript{1},
    Yingli Zhou\textsuperscript{3},
    Jieming Shi\textsuperscript{4},
    Chenhao Ma\textsuperscript{1}}
  \IEEEauthorblockA{\textsuperscript{1}The Chinese University of Hong Kong, Shenzhen, China\\
    \textsuperscript{2}Fudan University, Shanghai, China\\
    \textsuperscript{3}Lyon 1 University \& CNRS Liris, France\\
    \textsuperscript{4}The Hong Kong Polytechnic University, Hong Kong, China\\
    \{shulunzhang,yuweixu\}@link.cuhk.edu.cn, machenhao@cuhk.edu.cn\\
    lintingwang25@m.fudan.edu.cn, yingli.zhou@univ-lyon1.fr, jieming.shi@polyu.edu.hk}
  \thanks{Code is available at \url{https://github.com/alanzhang1001/GenTUS}.}
}
\hypersetup{pdfauthor={Shulun Zhang, Linting Wang, Yuwei Xu, Yingli Zhou, Jieming Shi, Chenhao Ma}}

\begin{document}
\maketitle

\begin{abstract}
Modern data lakes contain heterogeneous tables, and the information needed for a task is often scattered across many tables with different schemas, sources, and naming conventions.
Table union search (TUS) retrieves tables whose rows are semantically compatible with a query table for union, supporting data discovery, enrichment, and downstream analytics.
Existing TUS methods follow an encode--search--refine pipeline: they embed tables or columns, retrieve a candidate pool from a vector index, and refine it by query--candidate matching or reranking.
As a result, a unionable table missed by candidate retrieval cannot be recovered later, and searching and refining a pool for every query adds latency and storage costs that grow with the data lake.
We propose \method{}, a generative retrieval framework that reformulates TUS as constrained generation over discrete semantic table identifiers.
\method{} assigns each data-lake table a compact identifier learned under pairwise unionability supervision, which encourages unionable tables to share identifier prefixes, and trains a generator to produce the identifiers of unionable tables directly from the query table.
At query time, prefix-constrained beam search guarantees that every generated identifier maps to an existing table, and its output order serves directly as the ranking, with no candidate pool to search or rerank.
Experiments on seven public TUS benchmarks show that \method{} achieves the best overall retrieval quality, with the largest gains on the larger and more diverse benchmarks, while requiring substantially lower online retrieval time and retrieval-artifact storage than prior TUS methods.
\end{abstract}


\section{Introduction}
\label{sec:intro}

\begin{figure}[t]
    \centering
    \includegraphics[width=\linewidth]{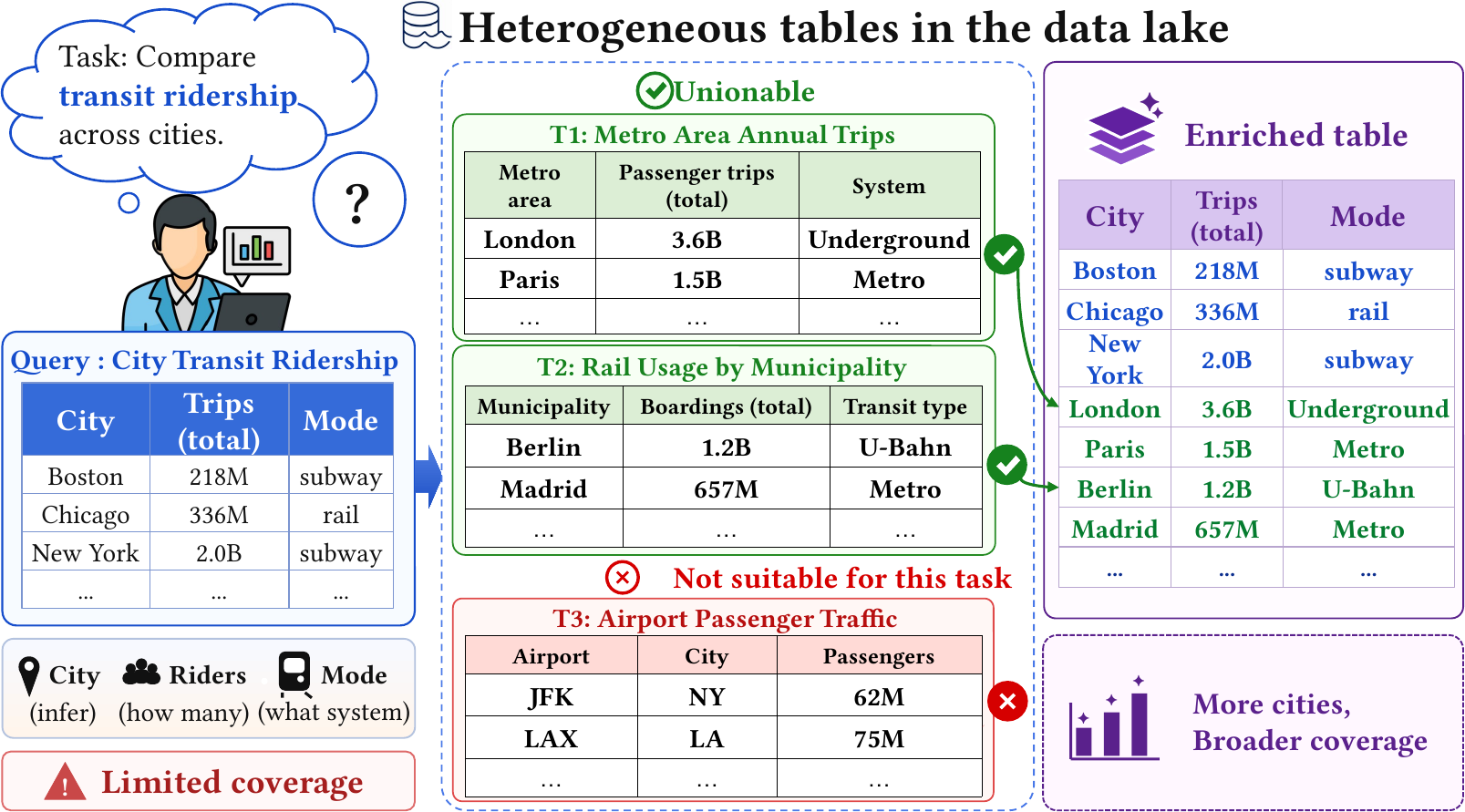}
    \caption{An example of table union search.}
    \label{fig:intro}
\end{figure}

Modern data lakes contain large collections of heterogeneous tables published by different organizations and curated under different schemas, metadata conventions, and quality standards~\cite{DBLP:journals/pvldb/NargesianZMPA19,8509315,DBLP:conf/sigmod/ZhangI20,10.14778/3750601.3750694,10.1145/2948674.2948675}.
To support downstream analysis, users often need to discover tables from a data lake that are relevant to a given analytical task.
The information needed for a task is often not contained in a single table; instead, it is scattered across multiple tables with overlapping semantics but different schemas, naming conventions, and coverage~\cite{10.1145/3299869.3300065,DBLP:journals/pvldb/DengCCYCYSWLCJZJZWYWT24,DBLP:journals/pvldb/NargesianZMPA19}.
To guide such discovery, a query table, which may come from an existing dataset, an intermediate analysis result, or a small seed table, provides a concrete specification of the desired attributes and semantics.
As illustrated in Fig.~\ref{fig:intro}, the goal is then to retrieve data-lake tables whose rows are semantically compatible with the query table for union.

This need motivates table union search (TUS), which retrieves tables from a data lake that can be unioned with a given query table. 
TUS is a key primitive for data discovery, integration, enrichment, exploratory analysis, and downstream analytics or model training~\cite{DBLP:journals/pvldb/NargesianZPM18,DBLP:journals/pacmmod/KhatiwadaFSCGMR23,chepurko2020arda,DBLP:journals/pvldb/CasteloRSBCF21}. 
At its core, TUS is a semantic compatibility problem: it seeks tables whose columns can be aligned with the query table so that their rows can be appended. 
Unlike keyword-based table search or join-oriented discovery, TUS cannot rely on keyword overlap or shared keys alone, because unionable tables may use different column names, have missing metadata, and share only sparse value overlap.

\begin{figure}[t]
    \centering
    \includegraphics[width=\linewidth]{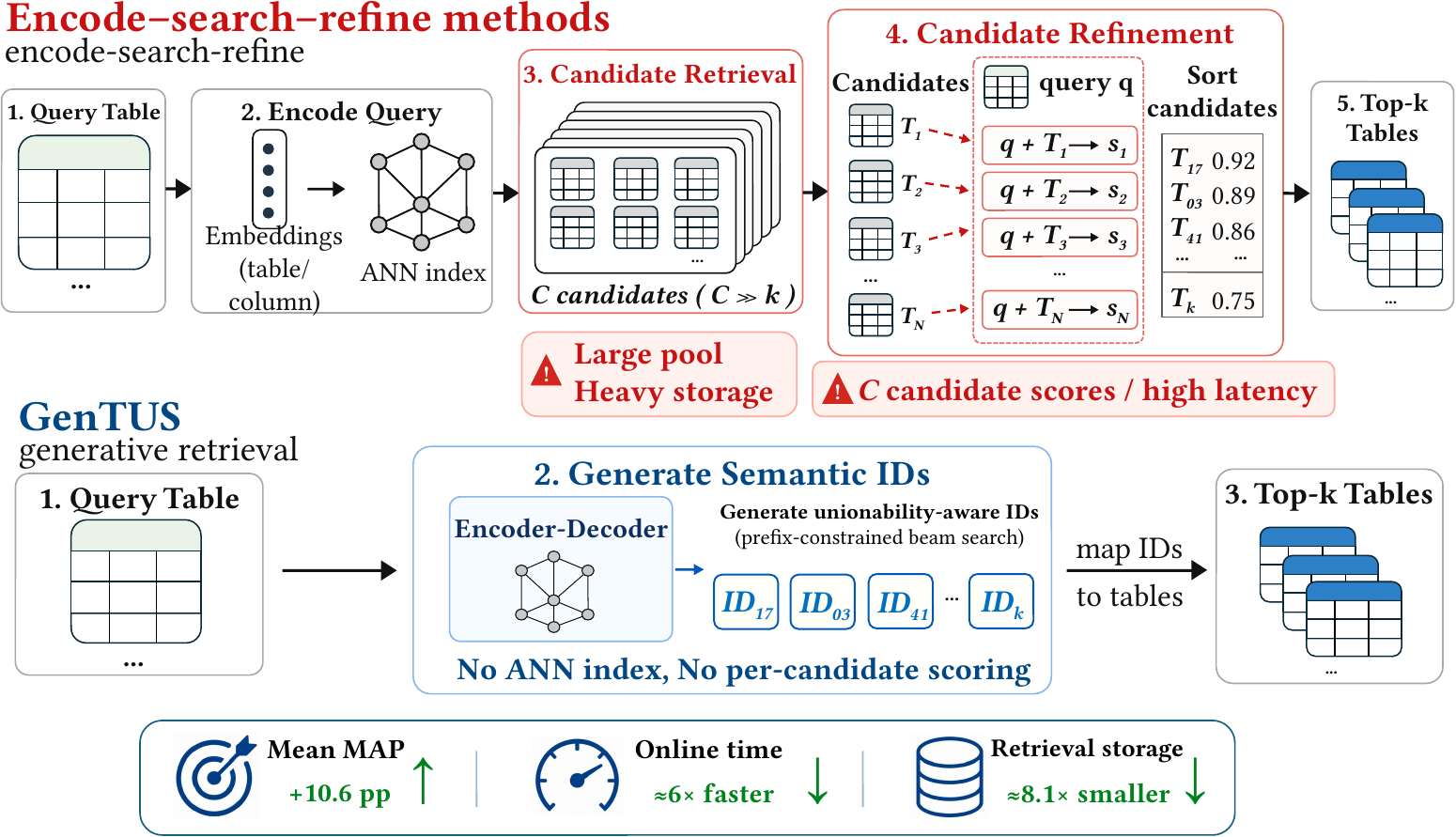}
    \caption{Prior TUS methods vs. GenTUS.}
    \label{fig:compare}
\end{figure}

{\bf Prior work.}
As summarized in Fig.~\ref{fig:compare}, existing TUS methods largely follow an \emph{encode--search--refine} pipeline with three stages:
{\it (1) encode}, where tables or their columns are encoded into dense representations and stored in a vector index;
{\it (2) search}, where the encoded query retrieves a pool of candidate tables through table-level search or column-level matches; and
{\it (3) refine}, where the candidates are rescored by a reranking model or a matching-based procedure to produce the final top-\(k\) unionable tables.
Methods differ mainly in how they encode and refine.
For example, Starmie~\cite{DBLP:journals/pvldb/FanWLZM23} and LIFTus~\cite{DBLP:conf/icde/QiuGTY25} encode each table at the column level and aggregate column-level evidence during retrieval or refinement to score candidate tables. 
In contrast, table-centric methods such as TACTUS~\cite{sun2026efficient} learn table-level embeddings for table-first candidate retrieval, followed by reranking with column-level evidence.

{\bf Motivation.}
In the conventional pipeline, a table must first be selected as a candidate before it can enter the final result.
Even an accurate matching procedure cannot recover a unionable table missed by the initial search.
Retrieving more candidates can reduce such misses, but it also leaves more tables to compare and rank.
The system must additionally store table or column representations and maintain the indexes used to search them.
As the data lake grows, maintaining retrieval quality can therefore require more work at query time, alongside substantial storage for candidate search.

Generative retrieval offers another possibility: a model can learn to predict the identifiers of relevant results directly from a query~\cite{DBLP:conf/nips/Tay00NBM000GSCM22}.
For TUS, each table can be assigned an identifier that serves as its address in the data lake.
Using examples of query tables and their unionable tables, a model can learn to generate the identifiers of the desired results.
When a new query arrives, the model predicts target identifiers, which are then mapped back to the corresponding tables.
The model learns which tables to return; the final lookup simply locates them.

This process avoids relying on an independent search stage to decide which tables can enter the final ranking.
It also removes the need for a separate round of comparing and reranking the retrieved candidates.
The lookup uses table identifiers rather than a dense vector-search index, offering a way to reduce the storage devoted to candidate retrieval.
Generative retrieval therefore provides a promising direction for addressing both the candidate-recall limitation and the retrieval costs of the conventional pipeline.

{\bf Our approach.}
Applying this idea to TUS raises two challenges.
First, the identifiers need to reflect which tables can be unioned.
Semantic identifiers can use shared prefixes to organize similar objects~\cite{DBLP:conf/nips/Tay00NBM000GSCM22,DBLP:conf/nips/RajputMSKVHHT0S23}, but similarity in table contents does not necessarily imply unionability.
Two tables may discuss the same topic while describing different attributes, making their records incompatible for union.
We therefore want unionable tables to share identifier prefixes. The model can then learn a common path to these tables, while the remaining tokens distinguish individual results.
Second, the model must generate identifiers assigned to actual tables.
An identifier that has never been assigned to a table cannot produce a retrieval result, so generation needs to stay within the set of existing identifiers.

We propose \textbf{\method{}}, a generative table union search framework that addresses these challenges through two designs (Fig.~\ref{fig:compare}).

{\bf Unionability-aware semantic identifier (SID) construction.}
Content-based quantization learns to reconstruct individual table representations, without using pairwise unionability annotations to guide code assignment.
We augment this training objective with a contrastive loss supervised by unionable table pairs.
The loss brings these tables closer in the quantizer's latent space, encouraging them to share coarse codes after quantization.
The generator can then learn common prefix paths to multiple unionable tables, while later tokens distinguish the individual results.

{\bf SID-tree-based constrained decoding.}
Without structural constraints, beam search can retain prefixes that cannot be completed into any existing table identifier.
Such prefixes consume beam capacity without contributing valid retrieval results.
We organize all valid SIDs into a prefix tree and restrict each decoding step to tokens that extend the current prefix along an edge of the tree.
This ensures that completed identifiers correspond to existing tables.
Beam search over the SID tree produces the ranked identifiers, which are then mapped to tables.

Experiments on seven public TUS benchmarks show that \method{} achieves the best overall retrieval quality, with the largest gains on the larger and more diverse benchmarks, while requiring lower retrieval-artifact construction time, online retrieval time, and retrieval-artifact storage than prior TUS methods.

In summary, our main contributions are as follows.
\begin{itemize}

\item We propose \method{}, a generative retrieval framework for table union search, reformulating it as constrained generation over the identifiers of existing lake tables.

\item We design unionability-aware identifier construction that incorporates
pairwise unionability supervision, so that unionable tables share identifier prefixes.

\item We use prefix-tree-based constrained decoding to ensure valid identifier generation and support efficient online retrieval.

\item We conduct experiments on seven public TUS benchmarks, demonstrating
\method{}'s advantages in retrieval quality and system efficiency over
state-of-the-art methods.

\end{itemize}

\smallskip\noindent\textbf{Outline.}
We introduce the preliminaries in Section~\ref{sec:prelim} and review related work in Section~\ref{sec:related}.
We present an overview of \method{} in Section~\ref{sec:overview}, and detail the offline and online phases in Sections~\ref{sec:offline} and~\ref{sec:online}.
Experimental results are reported in Section~\ref{sec:experiments}, and we conclude in Section~\ref{sec:conclusion}.

\section{Preliminaries}
\label{sec:prelim}

\label{sec:prelim:tus}

Let $\mathcal{D}=\{T_1,\ldots,T_N\}$ denote the candidate data lake, where each table $T_i$ consists of a set of columns $C_i$ and a collection of tuples defined over $C_i$.
Given a query table $q$, table union search aims to retrieve a ranked list of candidate tables from $\mathcal{D}$ that are unionable with $q$.
We operationalize unionability using each benchmark's released query--candidate annotations: for benchmark $b$, let $\mathcal{G}_b(q)$ be the set of candidates labeled unionable with query $q$. The criteria are benchmark-specific. Original TUS is attribute-based and permits union based on one or more columns, whereas SANTOS also considers semantic relationships between column pairs~\cite{DBLP:journals/pvldb/NargesianZPM18,DBLP:journals/pacmmod/KhatiwadaFSCGMR23}. GenTUS uses these table-level labels for training and evaluation. We write $\mu(q,T)$ for the score assigned to candidate $T$, with larger values indicating higher rank.
\begin{definition}[Top-$k$ Table Union Search]
\label{def:tus}
Following the standard TUS formulation~\cite{DBLP:journals/pvldb/NargesianZPM18,DBLP:journals/pacmmod/KhatiwadaFSCGMR23}, given a data lake $\mathcal{D}$, a query table $q$, and a positive integer $k$, top-$k$ table union search returns a ranked list of $k$ distinct tables $\pi_k(q,\mathcal{D})=(T_{(1)},\ldots,T_{(k)})$ ordered by a unionability score $\mu$: $\mu(q,T_{(j)})\ge\mu(q,T_{(j+1)})$ for $j=1,\ldots,k-1$.
\end{definition}

We partition the annotated pairs $\mathcal{G}\subseteq\mathcal{Q}\times\mathcal{D}$ into a training split $\mathcal{G}_{\mathrm{train}}\subseteq\mathcal{Q}_{\mathrm{train}}\times\mathcal{D}$ and a test split $\mathcal{G}_{\mathrm{test}}\subseteq\mathcal{Q}_{\mathrm{test}}\times\mathcal{D}$, where $\mathcal{Q}_{\mathrm{train}}\cap\mathcal{Q}_{\mathrm{test}}=\emptyset$; all annotation-dependent offline training signals use only $\mathcal{G}_{\mathrm{train}}$.

\subsection{Generative Retrieval}
\label{sec:prelim:gentus}

In this paper we study a \emph{generative} formulation of table union search.
Each candidate table $T_i\in\mathcal{D}$ is assigned a discrete \emph{semantic identifier} $s_i=(s_{i,1},\ldots,s_{i,\ell_i})$, with tokens drawn from a finite code vocabulary. In our implementation, $s_i$ consists of an $L$-token residual-quantized base code, optionally followed by a suffix token for collision resolution.
Let $\mathcal{S}_\mathcal{D}=\{s_i\mid T_i\in\mathcal{D}\}$ be the set of valid identifiers for the candidate data lake $\mathcal{D}$, and $\phi:\mathcal{S}_\mathcal{D}\to\mathcal{D}$ the bijective mapping from identifiers back to tables.
\begin{definition}[Generative TUS Retriever]
\label{def:gen-retriever}
A generative TUS retriever models the conditional probability $p_\theta(s\mid q)$ of an identifier given the query table.
At inference, it decodes identifiers from $\mathcal{S}_\mathcal{D}$ and returns the corresponding tables through the mapping $\phi$, producing the top-$k$ result of Definition~\ref{def:tus} in generation order.
\end{definition}

\section{Related Work}
\label{sec:related}

\subsection{Table Discovery and Table Union Search}
\label{sec:related:tus}

Table discovery covers keyword or dataset search~\cite{DBLP:conf/www/BrickleyBN19}, schema matching, joinable-table discovery~\cite{8509315,10.1145/3299869.3300065}, and table union search (TUS).
Early systems rely on metadata, schemas, value overlap, or semantic annotations~\cite{DBLP:conf/sigmod/YakoutGCC12,DBLP:conf/icde/BogatuFP020,koutras2021valentine,10.14778/1453856.1453916,10.14778/1920841.1921005}, which can be unreliable when metadata is incomplete or inconsistent~\cite{DBLP:journals/pvldb/NargesianZMPA19,DBLP:journals/pvldb/CasteloRSBCF21,10.14778/1920841.1921005}.

TUS retrieves tables that can extend a query table with additional rows under compatible column semantics~\cite{DBLP:journals/pvldb/NargesianZPM18,DBLP:journals/pvldb/NargesianZMPA19,DBLP:journals/pvldb/KhatiwadaSGM22,DBLP:journals/corr/abs-2301-04901,hu-etal-2023-automatic}.
Unlike keyword or joinable-table search, it concerns semantic compatibility rather than textual overlap or shared keys.
\textsc{DUST} further studies how to select diverse, novel tuples from unionable tables to enrich the query table~\cite{khatiwada2026dust}.

\textsc{LakeBench}~\cite{DBLP:journals/pvldb/DengCCYCYSWLCJZJZWYWT24} and \textsc{LakeCompass}~\cite{DBLP:journals/pvldb/ChaiDZCZCWZYWT24} support large-scale evaluation and end-to-end table discovery, respectively.
\textsc{TUNE} studies variation in human judgments of unionability~\cite{klimenkova2026tune}, while benchmark reevaluations examine how dataset characteristics affect retrieval performance~\cite{boutaleb-etal-2025-somethings}.

\subsection{Embedding-Based Table Union Search}
\label{sec:related:repr}

To handle schema and metadata heterogeneity, TUS methods learn table or column embeddings, building on table representation learning~\cite{DBLP:journals/jdiq/BadaroSP23}, including \textsc{TaBERT}~\cite{DBLP:conf/acl/YinNYR20}, \textsc{TaPas}~\cite{DBLP:conf/acl/HerzigNMPE20}, and \textsc{TURL}~\cite{10.1145/3542700.3542709}.
Related methods also use semantic types~\cite{hulsebos2019sherlockdeeplearningapproach,zhang2020satocontextualsemantictype}, relationship semantics~\cite{DBLP:journals/pacmmod/KhatiwadaFSCGMR23}, and schema-, value-, embedding-, or matching-based signals~\cite{DBLP:conf/icde/BogatuFP020,koutras2021valentine,DBLP:conf/icde/FernandezMNM19,dong2023deepjoinjoinabletablediscovery,DBLP:journals/pvldb/KayaliLFVOS24}.
\textsc{Starmie}~\cite{DBLP:journals/pvldb/FanWLZM23} and \textsc{LIFTus}~\cite{DBLP:conf/icde/QiuGTY25} use contextualized or multi-aspect column embeddings and aggregate column-level evidence to score tables.
\textsc{TACTUS}~\cite{sun2026efficient} retrieves candidates using table-level embeddings, followed by dual-evidence reranking with column-level signals.
\textsc{EasyTUS} introduces \textsc{TUSBench} and combines LLM-based representations with ANN search~\cite{otto2025easytus}, while \textsc{PGTUS} uses proximity-graph refinement and pruning for multi-vector retrieval~\cite{xie2025pgtus}.
These embedding-based methods typically follow an encode--search--refine pipeline, using indexes such as \textsc{HNSW}~\cite{DBLP:journals/pami/MalkovY20} to retrieve candidates before aggregation or reranking.
\method{} instead quantizes table embeddings into semantic identifiers and generates them directly under constrained decoding.

\subsection{Generative Retrieval}
\label{sec:related:gen}

Early differentiable search index and autoregressive retrieval methods formulate retrieval as generating candidate identifiers conditioned on a query~\cite{DBLP:conf/nips/Tay00NBM000GSCM22,DBLP:conf/nips/WangHWMWCXCZL0022,DBLP:conf/iclr/CaoI0P21,bevilacqua2022autoregressivesearchenginesgenerating}.
In recommendation, \textsc{TIGER} constructs residual-quantized item identifiers and trains a sequence model to generate relevant IDs~\cite{DBLP:conf/nips/RajputMSKVHHT0S23,DBLP:conf/nips/OordVK17}.
\textsc{LMIndexer} learns semantic IDs and a generative indexer through progressive training and self-supervision, while \textsc{FORGE} studies SID construction choices~\cite{jin2024lmindexer,fu2025forge}.
Other work improves tokenization, collaborative-semantic alignment, and ranking-oriented training~\cite{DBLP:conf/icde/ZhengHLCZCW24,DBLP:conf/cikm/0007BLZ0FNC24,DBLP:journals/tkde/QuFZL25,zhai2024actionsspeaklouderwords}.
\textsc{Birdie}~\cite{DBLP:journals/pvldb/GuoHMZGZ25} applies differentiable search indexes to natural-language-driven table discovery, using an encoder--decoder to unify indexing and search and strengthen query--table interaction.
\method{} focuses on table-to-table union search, where identifiers and generation scores must reflect table-pair unionability.
It shapes residual-quantized identifiers with pairwise unionability supervision and calibrates generation scores through a ranking objective.

\begin{figure*}[!tp]
\centering
\includegraphics[width=\linewidth]{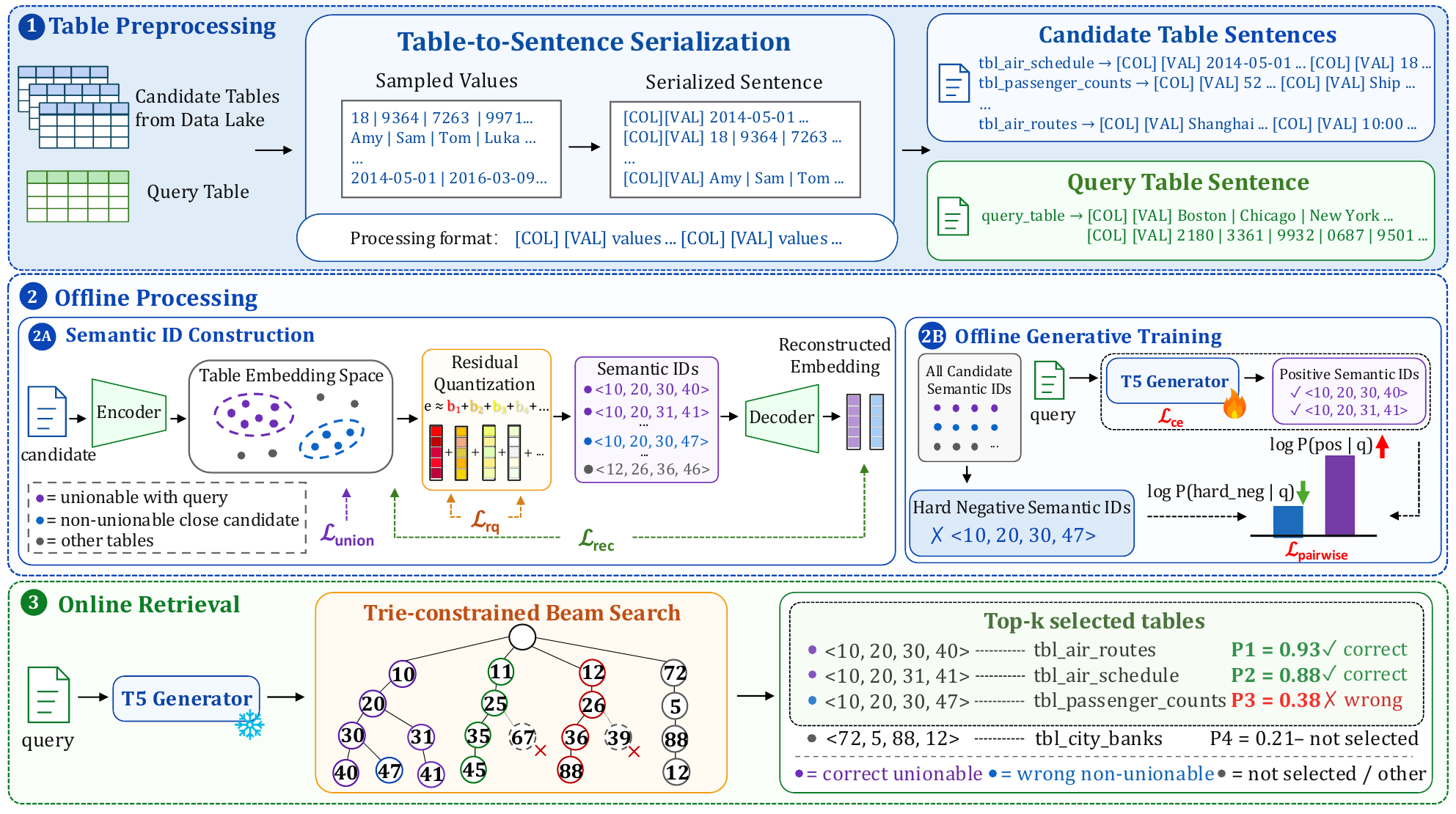}
\caption{Overview of \method{}.}
\label{fig:overview}
\end{figure*}

\section{Overview}
\label{sec:overview}

Fig.~\ref{fig:overview} illustrates the overall framework of \method{}.
\method{} reformulates table union search as constrained generation over semantic table identifiers.
Instead of maintaining a dense vector index and retrieving candidates by nearest-neighbor lookup, \method{} makes each candidate table addressable by a discrete semantic identifier.
Given a query table, an encoder--decoder generator directly produces identifiers of candidate tables that are likely to be unionable with the query, and the generated identifiers are mapped back to data-lake tables as retrieval results.

\textbf{Table preprocessing.}
\method{} serializes both candidate and query tables into table sentences that compactly capture sampled cell values and column-level structure, without using column names.
This shared format feeds the offline table representation step in Section~\ref{sec:offline:repr} and is reused for generator inputs during training and online retrieval.

\textbf{Offline processing.}
Given a data lake $\mathcal{D}$ and the training annotations $\mathcal{G}_{\mathrm{train}}$, \method{} constructs the generative search space in two parts.
The first part defines the discrete identifier space for candidate tables, and the second part trains a generator to search this space from a query table.

First, \method{} encodes candidate tables into table-level representations and performs unionability-aware semantic ID construction, as detailed in Section~\ref{sec:offline:repr} and Section~\ref{sec:offline:sid}.
For each candidate table $T \in \mathcal{D}$, \method{} feeds its serialized table sentence into a table encoder to obtain a table-level embedding, and then converts the embedding into a compact semantic identifier using a unionability-aware residual quantizer with learnable codebooks.
The quantizer is trained to preserve table content while shaping the discrete identifier space according to table-pair unionability, so that unionable tables are encouraged to share identifier prefixes.
This step produces the valid identifier set $\mathcal{S}_{\mathcal{D}}$, an identifier--table mapping $\phi$, and a prefix tree $\mathsf{Trie}_{\mathcal{D}}$ built over $\mathcal{S}_{\mathcal{D}}$ for constrained decoding.

Second, \method{} builds query-to-identifier training records and performs offline generative training, as detailed in Section~\ref{sec:offline:rec} and Section~\ref{sec:offline:gen}.
It trains an encoder--decoder generator $G_\theta$ over query-to-identifier records derived from $\mathcal{G}_{\mathrm{train}}$.
Each record takes a serialized training query table as input and uses the semantic identifier of an annotated unionable candidate as the target output sequence.
Besides the sequence generation objective, \method{} further uses mined hard-negative identifiers and a pairwise ranking objective to calibrate the relative scores of positive and negative identifiers.

\textbf{Online retrieval.}
Given a query table $q$, \method{} serializes $q$ using the same generator preprocessing procedure and feeds it into the trained generator, as described in Section~\ref{sec:online:retrieve}.
The decoder generates semantic identifiers token by token.
At each decoding step, the prefix tree $\mathsf{Trie}_{\mathcal{D}}$ restricts the next token to prefixes that can still lead to an existing table identifier in $\mathcal{S}_{\mathcal{D}}$.
This trie constraint prevents invalid identifiers from consuming beam-search capacity.
Completed identifiers are scored by the generator, mapped back to tables through $\phi$, deduplicated, and returned as the top-\(k\) unionable tables.
\section{Offline Processing}
\label{sec:offline}

The offline stage builds the generative candidate space used by \method{} and trains the retrieval generator.
It consists of two conceptual steps: Sections~\ref{sec:offline:repr}--\ref{sec:offline:sid} construct semantic identifiers for candidate tables, while Sections~\ref{sec:offline:rec}--\ref{sec:offline:gen} train a generator that maps query tables to identifiers of unionable candidates.

\subsection{Table Representation}
\label{sec:offline:repr}

\method{} serializes a table $T$ from its cell values, without column names. Each column $c_j$ contributes a bounded sequence of distinct non-empty values,
\[\sigma_j=[\mathrm{COL}]\circ[\mathrm{VAL}]\circ\operatorname{tok}(v_{j,1})\circ\cdots\circ\operatorname{tok}(v_{j,r_j}),\]
and the table sentence is $x_T=[\mathrm{CLS}]\circ\sigma_1\circ\cdots\circ\sigma_{|C_T|}$, where $\operatorname{tok}(\cdot)$ denotes tokenization and $\circ$ denotes concatenation.
The table encoder $F_\eta$ adapts the table-level encoder design of TACTUS~\cite{sun2026efficient}: a pretrained language model encodes $x_T$, the hidden states at the $[\mathrm{COL}]$ tokens represent the columns, and multi-head attention pooling with a learned query, followed by a projection and $\ell_2$ normalization, gives the table embedding $\mathbf{e}_T$.
This continuous table embedding $\mathbf{e}_T$ is then quantized into a discrete semantic identifier in the next step.

\subsection{Semantic Identifier Construction}
\label{sec:offline:sid}

\textbf{Design goals.}
\method{} represents each candidate table with a semantic identifier so that table retrieval can be formulated as sequence generation over a finite identifier space.
A useful identifier should satisfy two requirements: it should be {\em compact enough to be generated reliably by an autoregressive decoder}, and it should preserve the semantic structure needed for TUS so that {\em unionable tables share identifier prefixes}.

\begin{figure}[t]
    \centering
    \includegraphics[width=0.98\linewidth]{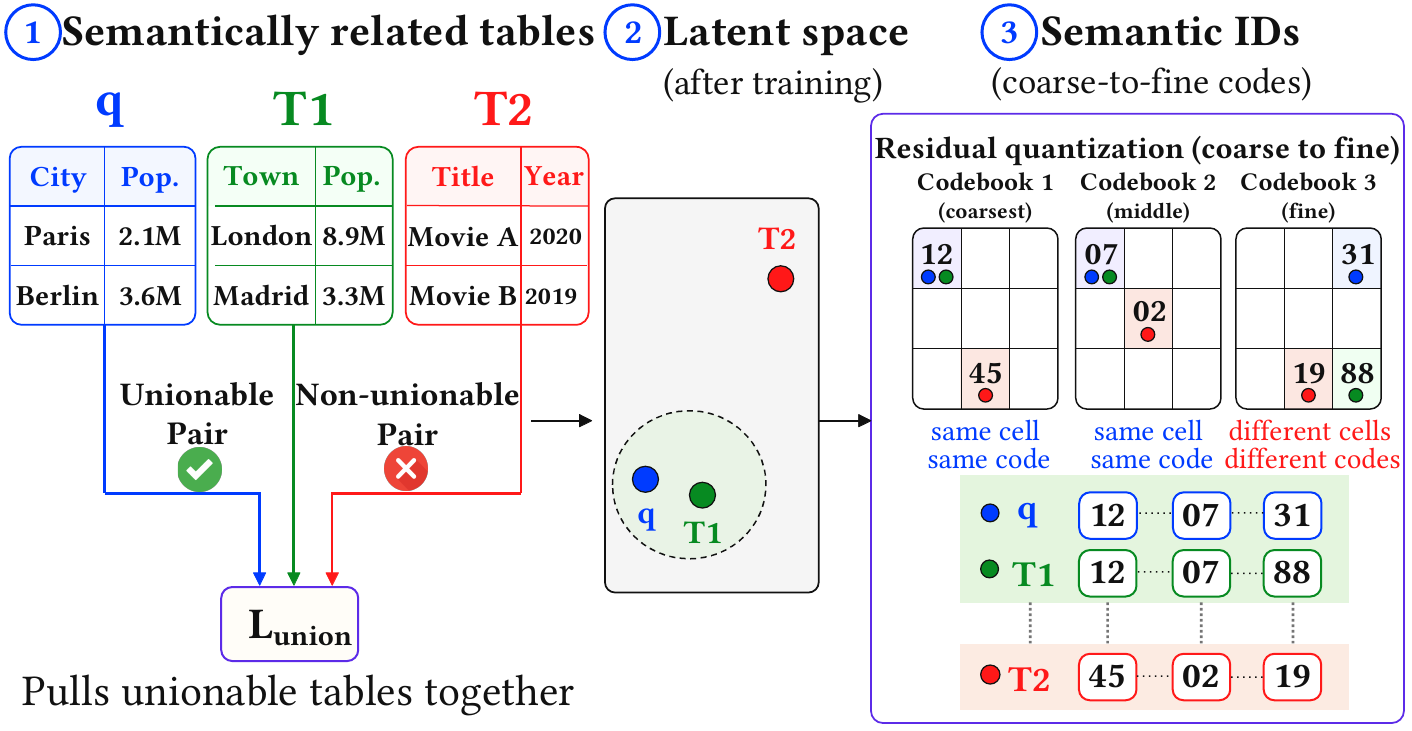}
    \caption{Unionability-aware semantic identifier construction. $\mathcal{L}_\mathrm{union}$ pulls unionable tables closer in the latent space, and residual quantization maps nearby tables to shared coarse codes with fine-grained distinctions.}
    \label{fig:sid}
\end{figure}

To this end, \method{} constructs identifiers with residual quantization.
Given the table embedding from Section~\ref{sec:offline:repr}, the quantizer maps each candidate table to a discrete sequence of code indices.
Because residual quantization represents a vector through multiple codebooks, the resulting identifier provides a compact coarse-to-fine code sequence that can be generated by the retrieval generator.
However, this quantization process by itself does not explicitly optimize identifiers for unionability.
\method{} therefore augments residual quantization with a unionability-aware contrastive objective, which pulls annotated unionable tables together in the quantizer latent space before discretization.
As illustrated in Fig.~\ref{fig:sid}, this objective and the residual quantizer jointly encourage unionable tables to share coarse codes while still allowing finer codes to distinguish individual tables.

\textbf{Residual quantization.}
The quantizer is fitted over candidate tables in $\mathcal{D}$, and its unionability term is supervised by $\mathcal{G}_{\mathrm{train}}$.
For each table $T_i\in\mathcal{D}$, we write $\mathbf{e}_i$ as shorthand for its table embedding $\mathbf{e}_{T_i}$ from Section~\ref{sec:offline:repr}.

The quantizer first maps this table embedding into a separate latent space used for code selection.
An MLP encoder $f_\omega$ produces the normalized quantizer latent vector
\[
\mathbf{z}_i
=
\frac{f_\omega(\mathbf{e}_i)}{\|f_\omega(\mathbf{e}_i)\|_2}.
\]
The latent vector is quantized by $L$ residual codebooks
$\{\mathcal{B}_\ell\}_{\ell=1}^L$,
where \(\mathcal{B}_\ell=\{\mathbf{b}_{\ell,1},\ldots,\mathbf{b}_{\ell,K}\}\).
Starting from $\mathbf{r}_i^{(0)}=\mathbf{z}_i$, the $\ell$-th code index is selected by nearest-neighbor lookup:
\[a_{i,\ell}=\arg\min_{k}\left\|\mathbf{r}_i^{(\ell-1)}-\mathbf{b}_{\ell,k}\right\|_2^2,\]
and the residual is updated as \(\mathbf{r}_i^{(\ell)}=\mathbf{r}_i^{(\ell-1)}-\mathbf{b}_{\ell,a_{i,\ell}}\).
The selected codewords are summed to form the quantized latent vector
\[
\widehat{\mathbf{z}}_i
=
\sum_{\ell=1}^{L}\mathbf{b}_{\ell,a_{i,\ell}}.
\]
The corresponding code indices form the base semantic identifier \(\bar{s}_i=(a_{i,1},\ldots,a_{i,L})\) of candidate table \(T_i\).

\textbf{Unionability-aware training and collision resolution.}
The quantizer is trained with three complementary objectives: a reconstruction loss, a residual quantization loss, and a unionability-aware contrastive loss.
For reconstruction, a decoder \(g'_\omega\) maps the quantized latent vector back to the table-embedding space as \(\widehat{\mathbf{e}}_i=g'_\omega(\widehat{\mathbf{z}}_i)\).
The reconstruction loss compares this reconstructed table embedding with the original table embedding:
\[
\mathcal{L}_\mathrm{rec}
=
\frac{1}{|\mathcal{D}|}
\sum_{T_i\in\mathcal{D}}
\left\|\mathbf{e}_i-\widehat{\mathbf{e}}_i\right\|_2^2.
\]

The residual quantization loss $\mathcal{L}_\mathrm{rq}$ trains the codebooks to represent the residual inputs at each level. Since nearest-neighbor codeword selection is non-differentiable, we follow the standard straight-through training of~\cite{DBLP:conf/nips/OordVK17,DBLP:conf/cvpr/LeeKKCH22} with a codebook term and a commitment term:
\[\mathcal{L}_\mathrm{rq}=\frac{1}{|\mathcal{D}|}\sum_{T_i\in\mathcal{D}}\sum_{\ell=1}^{L}\Big(\mathcal{L}^{(\ell)}_{\mathrm{codebook}}(i)+\beta\mathcal{L}^{(\ell)}_{\mathrm{commit}}(i)\Big).\]

\method{} further adds the unionability-aware contrastive objective over the latent vectors that pulls annotated unionable pairs together and contrasts them against other in-batch tables, aligning the latent space with TUS semantics~\cite{DBLP:conf/nips/KhoslaTWSTIMLK20}.
For contrastive training, we use only observed training query--candidate annotations whose query anchor is available in the benchmark corpus, and symmetrize these query--candidate pairs without inferring additional positive relations.
Let $\mathcal{P}$ be the resulting symmetrized set of positive pairs.

For a mini-batch \(B\), let \(\mathcal{P}_i^B=\{j\in B\setminus\{i\}:(i,j)\in\mathcal{P}\}\) be the positives of table \(T_i\) within the batch.
Mini-batches are sampled with positive-pair awareness so that annotated positive pairs are likely to appear in the same batch.
The unionability objective is
\[
\begin{aligned}
\mathcal{L}_\mathrm{union}
={}&-\frac{1}{|\mathcal{I}_B|}\sum_{i\in\mathcal{I}_B}\frac{1}{|\mathcal{P}_i^B|}\\
&\quad\times\sum_{j\in\mathcal{P}_i^B}
\log\frac{\exp(\mathbf{z}_i^\top\mathbf{z}_j/\tau)}
{\sum_{a\in B\setminus\{i\}}\exp(\mathbf{z}_i^\top\mathbf{z}_a/\tau)},
\end{aligned}
\]
where \(\mathcal{I}_B=\{i:\mathcal{P}_i^B\neq\emptyset\}\) and \(\tau\) is a temperature.

The full semantic-identifier training objective is
\[\mathcal{L}_\mathrm{SID}=\mathcal{L}_\mathrm{rec}+\mathcal{L}_\mathrm{rq}+
\lambda_u\mathcal{L}_\mathrm{union},\]
where $\mathcal{L}_\mathrm{rec}$ and $\mathcal{L}_\mathrm{rq}$ preserve the table-to-latent reconstruction and quantization behavior, while $\mathcal{L}_\mathrm{union}$ shapes the latent space so that unionable tables receive identifiers with shared coarse codes.

Since residual quantization maps a continuous space to a finite code space, multiple candidate tables may receive the same base identifier.
\method{} resolves such collisions by defining the final semantic identifier as
\[
\operatorname{sid}(T_i)=
\begin{cases}
\bar{s}_i, & \text{if } \bar{s}_i \text{ is unique},\\
\bar{s}_i\circ[\mathrm{SUF}_{h_i}], & \text{otherwise},
\end{cases}
\]
where $h_i$ is a unique suffix index assigned to $T_i$ within its collision group.
This makes each final identifier unique while preserving the shared base identifier for collided tables.
The valid identifier set is then $\mathcal{S}_{\mathcal{D}}=\{\operatorname{sid}(T):T\in\mathcal{D}\}$, with bijection $\phi:\mathcal{S}_{\mathcal{D}}\to\mathcal{D}$ used for generator training and online decoding.

\subsection{Training Instance Construction}
\label{sec:offline:rec}

After semantic identifiers are assigned, \method{} converts annotated query--candidate unionability pairs into identifier-level supervision for the retrieval generator.
This subsection constructs positive generation records and augments them with structure-preserving query views, so the generator learns to output identifiers of annotated unionable candidates under incidental layout variations.
It also builds indexing records that ground table contents to identifiers and hard-negative ranking records that provide pairwise ranking supervision.

Here, $\operatorname{ser}(\cdot)$ denotes the table-to-sentence serializer used to construct generator inputs.
We write
\(\mathcal{G}_{\mathrm{train}}(q)=\{T\in\mathcal{D}:(q,T)\in
\mathcal{G}_{\mathrm{train}}\}\) for the annotated positives of \(q\).

\textbf{Positive generation records.}
For each annotated pair \((q,T^+)\in\mathcal{G}_{\mathrm{train}}\),
\method{} serializes \(q\) as the generator input and uses the final
identifier of \(T^+\) as the target output sequence:
\[
(x_q,y^+)=(\operatorname{ser}(q),\operatorname{tok}(\operatorname{sid}(T^+))).
\]
Thus, each unionability annotation is rewritten as a sequence-to-sequence record from a query table to the identifier of a unionable candidate.

\textbf{Structural augmentation.}
\method{} constructs additional query views by sampling rows and varying column orders.
If $\operatorname{Aug}(q)$ denotes the set of augmented views of $q$, each view \(\widetilde{q}\in\operatorname{Aug}(q)\cup\{q\}\) keeps the same target identifier, yielding \((\operatorname{ser}(\widetilde{q}),y^+)\), where
\(y^+=\operatorname{tok}(\operatorname{sid}(T^+))\).
We use bounded augmentations so that each view preserves the main table semantics while varying incidental structure.
The resulting original and augmented positive records form
\(\mathcal{R}_{\mathrm{pos}}\).

\textbf{Indexing records.} \method{} also constructs self-mapping records that map a table's own content to its identifier:
\begin{equation*}
\bigl(
\operatorname{ser}(T),
\operatorname{tok}(\operatorname{sid}(T))
\bigr).
\end{equation*}
These records are created for candidate tables that appear as annotated positives in \(\mathcal{G}_{\mathrm{train}}\).
They ground semantic identifiers in table content, so the generator learns not only query-to-candidate supervision but also the basic content-to-identifier mapping.
The resulting records form the indexing set \(\mathcal{R}_{\mathrm{idx}}\).

\textbf{Hard-negative ranking records.}
Finally, \method{} constructs hard-negative records for the pairwise ranking objective.
For each training query $q$, it computes a lightweight table profile from sampled non-empty cell-value tokens.
Candidate tables in \(\mathcal{D}\setminus(\mathcal{G}_{\mathrm{train}}(q)\cup\{q\})\)
are ranked by Jaccard similarity to \(q\), and the top-\(n_{\mathrm{neg}}\) tables are used as mined hard negatives~\cite{xiong2020approximate}.
For \(\widetilde{q}\in\operatorname{Aug}(q)\cup\{q\}\), a ranking record has the form
\begin{equation*}
(\operatorname{ser}(\widetilde{q}), y^+, y^-)
=
\left(
\operatorname{ser}(\widetilde{q}),
\operatorname{tok}(\operatorname{sid}(T^+)),
\operatorname{tok}(\operatorname{sid}(T^-))
\right),
\end{equation*}
where \(T^+\in\mathcal{G}_{\mathrm{train}}(q)\) and
\(T^-\in\mathcal{D}\setminus(\mathcal{G}_{\mathrm{train}}(q)\cup\{q\})\).
These mined negatives share surface features with the query but are not annotated as positives, making them useful for calibrating generator scores in the ranking objective. The resulting triples form
\(\mathcal{R}_{\mathrm{pairwise}}\).

\subsection{Generator Training}
\label{sec:offline:gen}

Given the identifier-level records constructed in Section~\ref{sec:offline:rec}, \method{} trains a retrieval generator to map serialized query tables to semantic identifiers of unionable candidate tables.
The training records provide two complementary supervision signals.
Records in \(\mathcal{R}_{\mathrm{pos}}\cup\mathcal{R}_{\mathrm{idx}}\) are used with a token-level cross-entropy objective, which teaches the generator to produce valid target identifiers.
Records in \(\mathcal{R}_{\mathrm{pairwise}}\) are used with a pairwise ranking objective, which encourages annotated positive identifiers to receive higher sequence scores than mined hard-negative identifiers~\cite{DBLP:conf/emnlp/WisemanR16}.
This ranking objective operates over discrete identifier sequences and is distinct from $\mathcal{L}_\mathrm{union}$ in semantic identifier construction (Section~\ref{sec:offline:sid}), which shapes the continuous latent space before quantization.

\textbf{Generator parameterization.} 
\method{} instantiates the generator as a T5 encoder--decoder.
Since the decoder must generate semantic identifiers rather than only natural-language tokens, \method{} extends the tokenizer with identifier-specific tokens.
Specifically, the tokenizer is extended with $K$ code-index tokens shared across residual-code positions; the position of a token in the identifier sequence determines the corresponding residual codebook.
This lets the same code-index vocabulary represent each residual-code level while preserving the ordered structure of the identifier.
The tokenizer is also extended with suffix tokens for collision resolution and structural markers $[\mathrm{COL}]$ and $[\mathrm{VAL}]$ used by the generator serializer.
All other tokens are inherited from the base T5 tokenizer.

\textbf{Cross-entropy training.}
Given an input serialization $x$ and an identifier sequence $y$, where $y=\operatorname{tok}(\operatorname{sid}(T))$ for some candidate table $T$, the generator defines
\begin{equation*}
P_\theta(y\mid x)=
\prod_{t=1}^{|y|}
P_\theta(y_t\mid y_{<t},x).
\end{equation*}
For a positive target identifier $y^+$, the token-level cross-entropy loss is
\begin{equation*}
\mathcal{L}_\mathrm{ce}=
-\frac{1}{|y^+|}
\sum_{t=1}^{|y^+|}
\log P_\theta(y^+_t\mid y^+_{<t},x).
\end{equation*}
This loss is applied to positive generation records in \(\mathcal{R}_{\mathrm{pos}}\) and table-to-identifier indexing records in \(\mathcal{R}_{\mathrm{idx}}\).

\textbf{Pairwise ranking training.}
For ranking, \method{} scores an identifier sequence by its average conditional log-likelihood under the generator
\begin{equation*}
s_\theta(y\mid x)=
\frac{1}{|y|}
\sum_{t=1}^{|y|}
\log P_\theta(y_t\mid y_{<t},x).
\end{equation*}
Given a ranking triple $(x,y^+,y^-)$, where $y^+$ is the identifier of an annotated positive candidate and $y^-$ is the identifier of a mined hard-negative candidate, the pairwise ranking loss is
\begin{equation*}
\mathcal{L}_\mathrm{pairwise}=
\max\left(
0,
\gamma-
s_\theta(y^+\mid x)
+
s_\theta(y^-\mid x)
\right),
\end{equation*}
where $\gamma$ is the margin.
This loss is applied to hard-negative ranking records in \(\mathcal{R}_{\mathrm{pairwise}}\) and calibrates the generator so that positive identifiers receive higher average sequence log-likelihood than hard-negative identifiers.

The full generator objective is
\begin{equation*}
\mathcal{L}_\mathrm{gen}=
\mathcal{L}_\mathrm{ce}
+
\lambda_r
\mathcal{L}_\mathrm{pairwise}.
\end{equation*}
The cross-entropy term trains the generator to produce target identifiers, while the ranking term improves the relative ordering of positive and mined hard-negative identifiers under the model likelihood.
During online retrieval, \method{} does not perform a separate post-hoc query--candidate scoring step.
Valid identifiers returned by prefix-constrained beam search are ranked by the beam-search order induced by the generator scores.

\textbf{Offline pipeline summary.}
The pseudocode in Algorithm~\ref{alg:offline} summarizes how the preceding components form the complete offline pipeline.
Given the candidate lake, training queries, and unionability annotations, \method{} first encodes candidate tables and constructs valid semantic identifiers, then rewrites training annotations into identifier-level records and trains the retrieval generator.
The offline stage also produces the identifier mappings and prefix tree used by online constrained decoding.

\begin{algorithm}[t]
\caption{GenTUS Offline Training}
\label{alg:offline}
\KwIn{Candidate lake \(\mathcal{D}\), training queries \(\mathcal{Q}_{\mathrm{train}}\), annotations \(\mathcal{G}_{\mathrm{train}}\), table encoder \(F_\eta\), base generator \(G_\theta^{(0)}\)}
\KwOut{Quantizer \(Q_\omega\), trained generator \(G_\theta\), valid IDs \(\mathcal{S}_{\mathcal{D}}\), mappings \(\phi,\phi^{-1}\), prefix tree \(\mathsf{Trie}_{\mathcal{D}}\)}
\(E \leftarrow \{\mathbf{e}_{T_i}=F_\eta(T_i)\mid T_i\in\mathcal{D}\}\)\;
\(\mathcal{P}_{\mathrm{train}} \leftarrow \textsc{Symmetrize}(\{(q,T^+)\in\mathcal{G}_{\mathrm{train}}\mid q \in \mathcal{D}\})\)\;
\(\mathcal{L}_{\mathrm{SID}}\leftarrow\mathcal{L}_{\mathrm{rec}}+\mathcal{L}_{\mathrm{rq}}+\lambda_u\mathcal{L}_{\mathrm{union}}\)\;
\(Q_\omega \leftarrow \textsc{TrainUnionableAwareRQ}(E,\mathcal{P}_{\mathrm{train}};\mathcal{L}_{\mathrm{SID}})\)\;
Initialize \(\mathcal{I}_{\mathcal{D}} \leftarrow \emptyset\)\;
\ForEach{candidate table \(T_i\in\mathcal{D}\)}{
    \(\bar{s}_i \leftarrow \textsc{Quantize}(Q_\omega,\mathbf{e}_{T_i})\)\;
    \(\mathcal{I}_{\mathcal{D}} \leftarrow \mathcal{I}_{\mathcal{D}}\cup\{(T_i,\bar{s}_i)\}\)\;
}
\(\mathcal{I}_{\mathcal{D}} \leftarrow \textsc{ResolveCollisions}(\mathcal{I}_{\mathcal{D}},E)\)\;
\(\mathcal{S}_{\mathcal{D}}\leftarrow\{\operatorname{sid}(T_i)\mid T_i\in\mathcal{D}\}\)\;
\(\phi,\phi^{-1}\leftarrow\textsc{BuildMapping}(\mathcal{I}_{\mathcal{D}})\)\;
\(\mathcal{N} \leftarrow \textsc{MineHardNegatives}(\mathcal{Q}_{\mathrm{train}},\mathcal{D},\mathcal{G}_{\mathrm{train}})\)\;
\(\begin{aligned}[t]
& (\mathcal{R}_{\mathrm{pos}},\mathcal{R}_{\mathrm{idx}},\mathcal{R}_{\mathrm{pairwise}}) \leftarrow \\
& \quad\textsc{BuildTrainingRecords}(\mathcal{Q}_{\mathrm{train}},\mathcal{G}_{\mathrm{train}},\mathcal{D},\phi^{-1},\mathcal{N})
\end{aligned}\)\;
\(\mathcal{R}_{\mathrm{ce}}\leftarrow\mathcal{R}_{\mathrm{pos}}\cup\mathcal{R}_{\mathrm{idx}}\)\;
\(\mathcal{L}_{\mathrm{gen}}\leftarrow\mathcal{L}_{\mathrm{ce}}+\lambda_r\mathcal{L}_{\mathrm{pairwise}}\)\;
\(G_\theta\leftarrow\textsc{TrainGenerator}(G_\theta^{(0)},\mathcal{R}_{\mathrm{ce}},\mathcal{R}_{\mathrm{pairwise}};\mathcal{L}_{\mathrm{gen}})\)\;
\(\mathsf{Trie}_{\mathcal{D}} \leftarrow \textsc{BuildPrefixTree}(\{\operatorname{tok}(s):s \in \mathcal{S}_{\mathcal{D}}\})\)\;
\Return{\(Q_\omega,G_\theta,\mathcal{S}_{\mathcal{D}},\phi,\phi^{-1},\mathsf{Trie}_{\mathcal{D}}\)}\;
\end{algorithm}

\section{Online Retrieval}
\label{sec:online}

The online stage uses the generator and identifier structures produced offline to retrieve unionable tables through generation.
Given a query table, \method{} serializes it as the generator input, performs prefix-constrained decoding over valid semantic identifiers, and maps the completed identifiers back to candidate tables (Section~\ref{sec:online:retrieve}).
Thus, online retrieval is formulated as constrained identifier generation rather than dense-index lookup followed by explicit query--candidate reranking.

\subsection{Constrained Identifier Generation and Ranking}
\label{sec:online:retrieve}

Given a query table \(q\), \method{} applies the same serializer used during offline training, \(x_q=\operatorname{ser}(q)\).
The serialized query is fed into the trained encoder--decoder generator.
For an identifier sequence $s=(s_1,\ldots,s_{|s|})$, the generator defines the autoregressive probability
\[P_\theta(s\mid x_q)=\prod_{t=1}^{|s|}P_\theta(s_t\mid s_{<t},x_q).\]
Validity is enforced by the prefix-constrained decoding rule below.

\textbf{Prefix-constrained decoding.}
The prefix tree \(\mathsf{Trie}_{\mathcal{D}}\) is the online validity structure of \method{}. 
It stores the finite language of semantic identifiers for the candidate data lake $\mathcal{D}$ and constrains the decoder so that every completed sequence corresponds to a candidate table. This prevents beam search from spending capacity on invalid identifier sequences.

It is constructed over the tokenized final identifier set
$\{\operatorname{tok}(s):s\in\mathcal{S}_{\mathcal{D}}\}$.
For a decoded prefix \(y_{1:t}=(y_1,\ldots,y_t)\), \method{} defines the valid next-token set as
\[
\begin{aligned}
\mathcal{A}(y_{1:t})
=&\ \{v \mid y_{1:t}\circ v \text{ is a prefix in } \mathsf{Trie}_{\mathcal{D}}\} \\
&\ \cup
\{\mathrm{EOS} \mid y_{1:t}=\operatorname{tok}(s),\ s\in\mathcal{S}_{\mathcal{D}}\}.
\end{aligned}
\]
Continuation tokens are allowed only if they keep the decoded sequence on a valid identifier prefix, and \(\mathrm{EOS}\) is allowed whenever the current prefix already forms a complete valid identifier, even if that prefix also has valid suffix continuations.

At each decoding step, \method{} masks invalid next tokens before the beam update. 
If \(\ell_\theta(v\mid y_{1:t},x_q)\) denotes the generator logit (i.e., pre-softmax score) for token \(v\), the constrained logit is
\[
\ell_\theta^{\mathrm{valid}}(v\mid y_{1:t},x_q)=
\begin{cases}
\ell_\theta(v\mid y_{1:t},x_q), & v\in\mathcal{A}(y_{1:t}),\\
-\infty, & v\notin\mathcal{A}(y_{1:t}).
\end{cases}
\]
Beam search then expands only prefixes that remain valid under
\(\mathsf{Trie}_{\mathcal{D}}\).
The decoder returns a list of completed valid identifiers
\[\mathcal{Y}_q=[s^{(1)},\ldots,s^{(M)}],\]
where each \(s^{(m)}\in\mathcal{S}_{\mathcal{D}}\). 
\method{} preserves the beam-search order induced by the constrained generator scores as the identifier ranking.

\textbf{Table mapping and ranking.}
Prefix-constrained decoding returns an ordered list of valid identifiers
$\mathcal{Y}_q=[s^{(1)},\ldots,s^{(M)}]$, where each
$s^{(m)}\in\mathcal{S}_{\mathcal{D}}$.
\method{} maps these identifiers to candidate tables using
$\phi:\mathcal{S}_{\mathcal{D}}\to\mathcal{D}$:
\begin{equation*}
\widetilde{\mathcal{R}}_q
=
[\phi(s^{(1)}),\ldots,\phi(s^{(M)})].
\end{equation*}
If duplicate tables appear after mapping, \method{} keeps the earliest occurrence
and removes later duplicates.
Let $\mathcal{R}_q$ denote the resulting deduplicated ordered list.
The final top-$k$ result is
\begin{equation*}
\pi_k(q,\mathcal{D})
=
\operatorname{FirstK}_k(\mathcal{R}_q).
\end{equation*}
The order of $\mathcal{R}_q$ is inherited from the constrained beam search, so
\method{} uses it directly as the retrieval ranking without post-hoc
query--candidate alignment scoring over an ANN-retrieved candidate pool.
The pseudocode in Algorithm~\ref{alg:online} summarizes the online retrieval pipeline.

\begin{algorithm}[t]
\caption{GenTUS Online Generation and Retrieval}
\label{alg:online}
\KwIn{Query table \(q\), generator \(G_\theta\), mapping \(\phi\), prefix tree \(\mathsf{Trie}_{\mathcal{D}}\), result size \(k\)}
\KwOut{Top-\(k\) unionable tables \(\pi_k(q,\mathcal{D})\)}
\(x_q \leftarrow \operatorname{ser}(q)\)\;
\(\mathcal{Y}_q\leftarrow\textsc{PrefixConstrainedBeamSearch}(G_\theta,x_q,\mathsf{Trie}_{\mathcal{D}})\)\;
Initialize \(\mathcal{R}_q\leftarrow[\,]\)\;
\ForEach{\(s\in\mathcal{Y}_q\)}{
    \(T\leftarrow\phi(s)\)\;
    \If{\(T\notin \mathcal{R}_q\)}{
        Append \(T\) to \(\mathcal{R}_q\)\;
    }
}
\Return{\(\operatorname{FirstK}_k(\mathcal{R}_q)\)}\;
\end{algorithm}

\section{Experiments}
\label{sec:experiments}

\newcommand{\stdcell}[1]{$\pm$#1}

\begin{table}[t]
\centering
\setlength{\tabcolsep}{2pt}
\caption{Dataset statistics over data lake tables.}
\label{tab:datasets}
\begin{tabular}{lrrrrrrr}
\toprule
Statistic & TUS-S & TUS-L & SAN-S & SAN-L & Wiki & LB-1K & LB-30K \\
\midrule
\#Tables     & 1{,}530  & 5{,}043  & 550    & 11{,}090  & 40{,}752  & 7{,}970   & 7{,}970 \\
\#Cols       & 14{,}810 & 54{,}923 & 6{,}322 & 123{,}477 & 106{,}744 & 151{,}572 & 151{,}572 \\
Avg.\ \#Cols & 9.7      & 10.9     & 11.5   & 11.1      & 2.6       & 19.0      & 19.0 \\
Avg.\ \#Rows & 4{,}466  & 1{,}915  & 6{,}921 & 7{,}675   & 51        & 936      & 19{,}138 \\
\bottomrule
\end{tabular}
\end{table}

We evaluate \method{} through four research questions.
\begin{itemize}
\item \textbf{RQ1 (Retrieval quality).} Does \method{} match or outperform state-of-the-art table union search methods on standard TUS benchmarks (Section~\ref{sec:exp:quality})?
\item \textbf{RQ2 (Efficiency and storage).} Does \method{} lower retrieval-artifact construction time, online retrieval time, and index storage (Section~\ref{sec:exp:efficiency})?
\item \textbf{RQ3 (Robustness to candidate-space size).} How does \method{}'s retrieval quality vary across candidate subsets of different sizes from the same data lake (Section~\ref{sec:exp:robustness})?
\item \textbf{RQ4 (Ablation and analysis).} How do \method{}'s key components and decoding choices affect retrieval quality and the accuracy--latency trade-off (Section~\ref{sec:exp:ablation})?
\end{itemize}

\begin{table*}[t]
\centering
\caption{Retrieval quality at the cutoff $k$ shown for each dataset, reporting MAP, P, and R at $k$. \method{} reports means over three runs, with standard deviations in the final row. In each column, the best result is in bold on a darker background, and the second-best is on a lighter one.}
\label{tab:main}
\fontsize{7.5}{8.8}\selectfont
\setlength{\tabcolsep}{1.3pt}
\renewcommand{\arraystretch}{1.17}
\newcommand{\resultheader}[1]{\textbf{#1}}
\newcommand{\resultgroup}[1]{\multicolumn{3}{c}{\resultheader{#1}}}
\newcommand{\resultmetrics}{\resultheader{MAP$\uparrow$} & \resultheader{P$\uparrow$} & \resultheader{R$\uparrow$}}
\arrayrulecolor{resultdivider}
\setlength{\arrayrulewidth}{0.3pt}
\begin{tabular*}{\textwidth}{@{\extracolsep{\fill}}l*{7}{|ccc}@{}}
\arrayrulecolor{black}\specialrule{0.8pt}{0pt}{0pt}\arrayrulecolor{resultdivider}
\rowcolor{resulthead}
\resultheader{Method} & \resultgroup{TUS-S ($k{=}60$)} & \resultgroup{TUS-L ($k{=}60$)} & \resultgroup{SAN-S ($k{=}10$)} & \resultgroup{SAN-L ($k{=}20$)} & \resultgroup{Wiki ($k{=}40$)} & \resultgroup{LB-1K ($k{=}60$)} & \resultgroup{LB-30K ($k{=}60$)} \\
\rowcolor{resultsubhead}
 & \resultmetrics & \resultmetrics & \resultmetrics & \resultmetrics & \resultmetrics & \resultmetrics & \resultmetrics \\
\arrayrulecolor{black}\specialrule{0.5pt}{0pt}{0pt}\arrayrulecolor{resultdivider}
Sherlock & 0.970 & 0.974 & 0.333 & 0.978 & 0.982 & 0.234 & 0.874 & 0.884 & 0.713 & 0.301 & 0.304 & 0.414 & 0.105 & 0.187 & 0.157 & 0.368 & 0.164 & 0.595 & 0.364 & 0.161 & 0.588 \\
SATO & 0.976 & 0.979 & 0.335 & 0.983 & 0.986 & 0.236 & 0.850 & 0.870 & 0.701 & 0.331 & 0.339 & 0.449 & 0.160 & 0.224 & 0.192 & 0.364 & 0.165 & 0.596 & 0.373 & 0.163 & 0.591 \\
\specialrule{0.3pt}{0pt}{0pt}
{SANTOS} & {0.865} & {0.867} & {0.296} & {0.970} & {0.978} & {0.235} & {0.900} & {0.904} & {0.733} & \cellcolor{secondfill}{0.512} & \cellcolor{secondfill}{0.485} & \cellcolor{secondfill}{0.632} & {0.160} & {0.305} & {0.243} & {0.242} & {0.096} & {0.360} & {0.243} & {0.096} & {0.361} \\
\specialrule{0.3pt}{0pt}{0pt}
Starmie & 0.939 & 0.951 & 0.324 & 0.932 & 0.940 & 0.221 & \cellcolor{secondfill}0.972 & \cellcolor{secondfill}0.974 & 0.791 & 0.388 & 0.387 & 0.503 & 0.247 & 0.396 & 0.324 & 0.304 & 0.088 & 0.335 & 0.297 & 0.084 & 0.324 \\
LIFTus & 0.961 & 0.967 & 0.330 & 0.989 & 0.992 & \cellcolor{secondfill}0.237 & 0.939 & 0.932 & 0.697 & 0.487 & 0.352 & 0.437 & 0.082 & 0.174 & 0.144 & 0.107 & 0.054 & 0.193 & 0.109 & 0.053 & 0.191 \\
TACTUS & \cellcolor{secondfill}{0.992} & {0.994} & \cellcolor{secondfill}{0.341} & \cellcolor{bestfill}\textbf{{0.998}} & \cellcolor{bestfill}\textbf{{0.998}} & \cellcolor{bestfill}\textbf{{0.238}} & \cellcolor{secondfill}0.972 & 0.972 & \cellcolor{secondfill}0.792 & {0.469} & {0.439} & {0.575} & 0.432 & 0.554 & 0.455 & {0.409} & {0.156} & {0.526} & {0.417} & {0.159} & {0.540} \\
\specialrule{0.3pt}{0pt}{0pt}
{Sup-BiEncoder} & {0.991} & \cellcolor{secondfill}{0.995} & \cellcolor{secondfill}{0.341} & {0.993} & \cellcolor{secondfill}{0.994} & \cellcolor{secondfill}{0.237} & {0.857} & {0.870} & {0.707} & {0.300} & {0.336} & {0.401} & {0.252} & {0.426} & {0.368} & \cellcolor{secondfill}{0.449} & \cellcolor{secondfill}{0.200} & \cellcolor{secondfill}{0.683} & {0.447} & {0.201} & {0.682} \\
\specialrule{0.3pt}{0pt}{0pt}
{BIRDIE} & {0.927} & {0.942} & {0.330} & {0.822} & {0.856} & {0.212} & {0.416} & {0.490} & {0.403} & {0.139} & {0.161} & {0.259} & \cellcolor{secondfill}{0.625} & \cellcolor{secondfill}{0.683} & \cellcolor{secondfill}{0.570} & {0.430} & {0.168} & {0.555} & \cellcolor{secondfill}{0.524} & \cellcolor{secondfill}{0.212} & \cellcolor{secondfill}{0.685} \\
\arrayrulecolor{black}\specialrule{0.5pt}{0pt}{0pt}
\textbf{GenTUS} & \cellcolor{bestfill}\textbf{0.997} & \cellcolor{bestfill}\textbf{0.997} & \cellcolor{bestfill}\textbf{0.342} & \cellcolor{secondfill}0.997 & \cellcolor{bestfill}\textbf{0.998} & \cellcolor{bestfill}\textbf{0.238} & \cellcolor{bestfill}\textbf{0.988} & \cellcolor{bestfill}\textbf{0.989} & \cellcolor{bestfill}\textbf{0.804} & \cellcolor{bestfill}\textbf{0.532} & \cellcolor{bestfill}\textbf{0.558} & \cellcolor{bestfill}\textbf{0.666} & \cellcolor{bestfill}\textbf{0.704} & \cellcolor{bestfill}\textbf{0.773} & \cellcolor{bestfill}\textbf{0.638} & \cellcolor{bestfill}\textbf{0.602} & \cellcolor{bestfill}\textbf{0.238} & \cellcolor{bestfill}\textbf{0.766} & \cellcolor{bestfill}\textbf{0.613} & \cellcolor{bestfill}\textbf{0.240} & \cellcolor{bestfill}\textbf{0.765} \\
\textit{Std. ($\pm$)} & .0032 & .0027 & .0008 & .0007 & .0004 & .0001 & .0087 & .0077 & .0073 & .0268 & .0115 & .0129 & .0427 & .0387 & .0243 & .0228 & .0031 & .0222 & .0229 & .0050 & .0044 \\
\bottomrule
\end{tabular*}

\arrayrulecolor{black}
\end{table*}

\subsection{Experimental Setup}
\label{sec:exp:setup}

\textbf{Datasets.}
We evaluate on seven TUS datasets, summarized in Table~\ref{tab:datasets}:
\textsf{TUS\_small}, \textsf{TUS\_large}, \textsf{SANTOS\_small}, \textsf{SANTOS\_large}, \textsf{wiki\_union}, and two LakeBench-derived OpenData row-cap variants~\cite{DBLP:journals/pvldb/DengCCYCYSWLCJZJZWYWT24}.
The two LakeBench variants share the same tables, queries, and ground truth, but cap each CSV table at $1{,}000$ and $30{,}000$ rows, respectively~\cite{DBLP:journals/pvldb/DengCCYCYSWLCJZJZWYWT24}.
Positive labels follow each benchmark's native protocol: TUS is attribute-based, SANTOS is relationship-aware, and wiki\_union and LakeBench use their released annotations.
Together, the datasets cover a range of sizes, domains, and table-content scales.
All methods use $k=60$ for \textsf{TUS} and LakeBench, $k=10$ for \textsf{SANTOS\_small}, $k=20$ for \textsf{SANTOS\_large}, and $k=40$ for \textsf{wiki\_union}. For LakeBench, this differs from the original OpenData setting of $k=50$~\cite{DBLP:journals/pvldb/DengCCYCYSWLCJZJZWYWT24}.

\textbf{Protocol.}
Following the partition in Section~\ref{sec:prelim:tus}, we split the query tables of each benchmark into disjoint training and test queries at an 8:2 ratio.
All training objectives (i.e., semantic identifier construction ($\mathcal{L}_{\mathrm{union}}$), generator training, and hard-negative mining) use only $\mathcal{G}_{\mathrm{train}}$.
At inference, each test query is decoded against $\mathcal{S}_{\mathcal{D}}$.
For \method{}, we repeat the main retrieval-quality experiment three times with different random seeds and report the mean and standard deviation in Table~\ref{tab:main}.

\textbf{Baselines.}
We compare against six prior TUS methods and two reference baselines. The prior TUS methods fall into three groups: the semantic-type-based \textsf{Sherlock}~\cite{hulsebos2019sherlockdeeplearningapproach} and \textsf{SATO}~\cite{zhang2020satocontextualsemantictype}; the relationship-based \textsf{SANTOS}~\cite{DBLP:journals/pacmmod/KhatiwadaFSCGMR23}; and the representation-based \textsf{Starmie}~\cite{DBLP:journals/pvldb/FanWLZM23} and \textsf{LIFTus}~\cite{DBLP:conf/icde/QiuGTY25}, which are column-centric, and \textsf{TACTUS}~\cite{sun2026efficient}, which is table-centric. The two reference baselines are not existing TUS methods; we include them as controls. \textsf{Sup-BiEncoder}, which we implement, controls for supervision: it trains a BERT dual encoder on the same training annotations $\mathcal{G}_{\mathrm{train}}$ that \method{} uses and ranks candidates directly by embedding similarity. \textsf{BIRDIE}~\cite{DBLP:journals/pvldb/GuoHMZGZ25} controls for the generative paradigm: it is a generative retriever designed for natural-language table discovery, which we adapt to take serialized query tables as queries and train on the same annotations $\mathcal{G}_{\mathrm{train}}$.

We run the authors' public implementations of all published baselines and keep each method's recommended or best-reported configuration.
For retrieval baselines, this includes the published candidate-space size, ANN/HNSW parameters, and reranking budget when applicable.

\textbf{Metrics.}
Following prior TUS work~\cite{DBLP:journals/pvldb/NargesianZPM18,DBLP:journals/pacmmod/KhatiwadaFSCGMR23,DBLP:journals/pvldb/FanWLZM23,DBLP:journals/pvldb/KhatiwadaSGM22}, we report Mean Average Precision at $k$ (MAP@$k$), Precision at $k$ (P@$k$), and Recall at $k$ (R@$k$).
For benchmark $b$ and query table $q$, let $\mathcal{G}_b(q)$ be the released set of candidates labeled unionable with $q$, and let $\mathcal{R}_k(q)$ be the ranked top-$k$ results. We define $\mathrm{P}@k=\frac{|\mathcal{R}_k(q)\cap\mathcal{G}_b(q)|}{k}$ and $\mathrm{R}@k=\frac{|\mathcal{R}_k(q)\cap\mathcal{G}_b(q)|}{|\mathcal{G}_b(q)|}$. For AP@$k$, each labeled candidate $g\in\mathcal{G}_b(q)$ contributes $\mathrm{P}@r_q(g)$ if it appears at rank $r_q(g)\leq k$, and $0$ otherwise; AP@$k$ divides the sum of these contributions by $\min(k,|\mathcal{G}_b(q)|)$, and MAP@$k$ averages AP@$k$ over queries.

Since a query has $|\mathcal{G}_b(q)|$ labeled positives, $\mathrm{P}@k\le|\mathcal{G}_b(q)|/k$ and $\mathrm{R}@k\le k/|\mathcal{G}_b(q)|$: R@$k$ is low for all methods on TUS, whose queries have more than $k$ positives, and P@$k$ is low on LakeBench, whose queries have fewer.

For a fixed data lake, retrieval-artifact construction time covers computing representations from preloaded candidate tables and building retrieval structures.
Online retrieval time measures the time spent processing the full query workload for each dataset.
Retrieval-artifact storage includes candidate-specific representations, identifiers, indexes, and retained caches; fixed model weights and shared knowledge-base assets are excluded. Storage is reported in MiB ($2^{20}$ bytes).

\textbf{Implementation.}
The table encoder is \textsf{BERT-base}~\cite{devlin2019bertpretrainingdeepbidirectional}. 
To align with prior TUS work that treats column names as missing or unreliable metadata in open data lakes, each table is serialized without column names, using all columns and up to 24 sampled values per column under a 512-token budget.
Multi-head attention then pools the resulting column representations into one table vector.
The semantic identifier uses a residual quantizer with $L$ codebooks of size $K$, with suffix tokens used only to disambiguate duplicate base codes.
The quantizer is trained with a unionability-aware contrastive term of weight $\lambda_u{=}0.1$ and temperature $\tau{=}0.1$.
The generator is a \textsf{T5-base}~\cite{DBLP:journals/jmlr/RaffelSRLNMZLL20} encoder--decoder. We train it with cross-entropy and the pairwise ranking objective, using margin $\gamma{=}0.2$ and weight $\lambda_r{=}0.2$.
We train the generator for 30 or 80 epochs, depending on the dataset. The accompanying code repository lists $L$, $K$, and the training hyperparameters for each dataset.
Online decoding uses prefix-constrained beam search with beam width $\max(2k,100)$. All experiments run on 8 NVIDIA A100-80GB GPUs.

\begingroup
\setlength{\textfloatsep}{14pt plus 2pt minus 2pt}
\setlength{\dbltextfloatsep}{14pt plus 2pt minus 2pt}
\begin{figure*}[!t]
    \centering
    \includegraphics[width=0.98\linewidth]{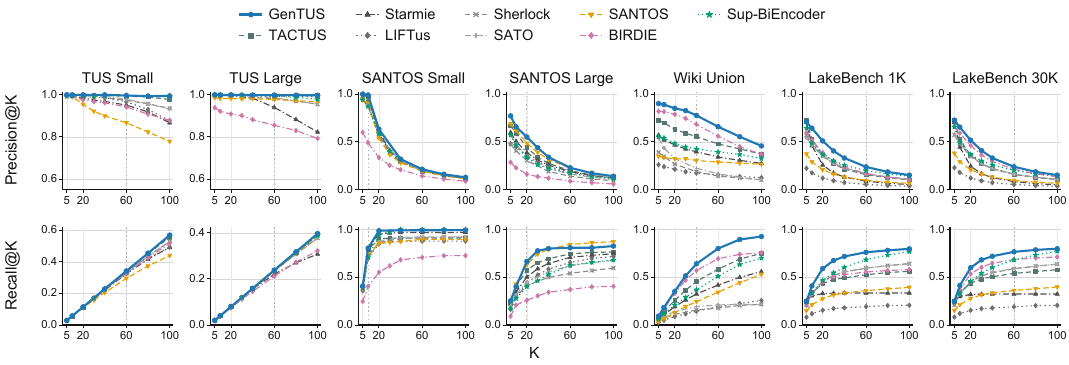}
    \caption{Precision@$k$ and Recall@$k$ as $k$ increases on the seven datasets. \method{} maintains stronger precision--recall trade-offs across different $k$ values.}
    \label{fig:e2}
\end{figure*}

\subsection{Retrieval Quality}
\label{sec:exp:quality}

Table~\ref{tab:main} reports MAP, P, and R at each dataset's evaluation cutoff $k$. For \method{}, we additionally report the mean and standard deviation over three independent runs.
\method{} leads overall, with the best result on most dataset--metric pairs.
It also achieves higher MAP than the adapted generative baseline \textsf{BIRDIE} on all seven datasets.

\textbf{Near-ceiling benchmarks.}
On \textsf{TUS\_small}, \textsf{TUS\_large}, and \textsf{SANTOS\_small}, most methods are already close to the ceiling, so the margins are small.
On \textsf{TUS\_small}, \method{} reaches MAP $99.7\%$, P $99.7\%$, and R $34.2\%$, while the strongest baselines reach $99.2\%$ MAP and $99.5\%$ P.
On \textsf{TUS\_large}, \method{} and \textsf{TACTUS} achieve nearly identical retrieval quality.
These lakes are semantically clean and small enough that most methods perform well, and \method{} remains best or tied-best on most of their metrics.

\textbf{Larger and more diverse benchmarks.}
The advantage widens on \textsf{SANTOS\_large}, \textsf{wiki\_union}, and the two \textsf{LakeBench} splits, where larger scale and greater semantic diversity make retrieval more challenging.
On \textsf{SANTOS\_large}, \method{} outperforms \textsf{SANTOS} and \textsf{TACTUS} across all three metrics.
On \textsf{wiki\_union}, \textsf{BIRDIE} remains competitive, but \method{} achieves higher precision and recall.
On \textsf{LakeBench\_1k}, \method{} achieves $60.2\%$ MAP, $15.3$ percentage points above \textsf{Sup-BiEncoder}, the strongest baseline. On \textsf{LakeBench\_30k}, it leads the strongest baseline, \textsf{BIRDIE}, by $8.9$ points.

On the larger and more diverse lakes, the baselines are more affected by their respective retrieval bottlenecks: type-based methods rely on coarse column categories, column-aggregation methods search over a much larger column space, and table-vector pipelines compress table evidence before retrieving a bounded candidate pool.
\method{} instead directly generates unionability-supervised semantic identifiers, preserving unionability signals through retrieval.

\textbf{Varying \(k\).}
Fig.~\ref{fig:e2} reports Precision@$k$ and Recall@$k$ as $k$ grows.
As expected, precision generally decreases as $k$ grows.
Across most evaluated $k$ values, \method{} provides the strongest overall precision--recall trade-off.
On \textsf{TUS\_small} and \textsf{TUS\_large}, \method{} keeps precision near $0.99$ through $k{=}100$; \textsf{TACTUS} and \textsf{Sup-BiEncoder} also retain high precision, and \textsf{Sup-BiEncoder} reaches slightly lower recall than \method{} on \textsf{TUS\_large} at larger $k$.
On \textsf{SANTOS\_large}, \textsf{wiki\_union}, and the two \textsf{LakeBench} splits, \method{} generally maintains higher precision as recall increases with $k$.
The varying-$k$ comparison therefore confirms \method{}'s overall advantage across operating points while also showing where individual baselines, such as \textsf{TACTUS} and \textsf{Sup-BiEncoder} on the two \textsf{TUS} benchmarks, remain competitive.

\begin{figure*}[t]
    \centering
    \includegraphics[width=\textwidth]{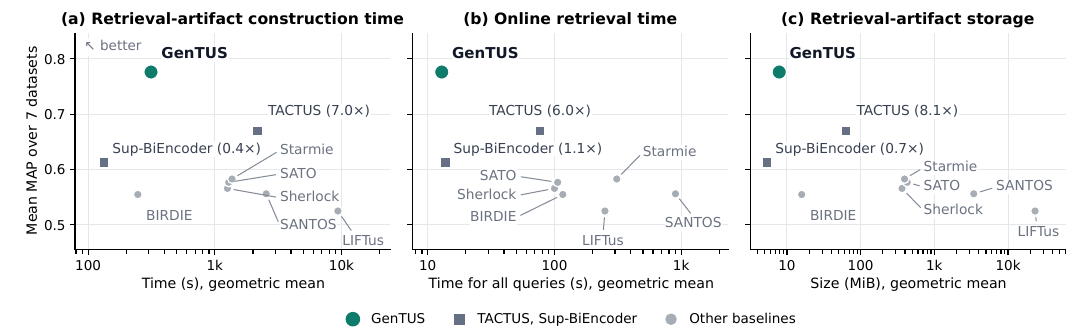}
    \caption{Retrieval quality versus (a) retrieval-artifact construction time, (b) online retrieval time, and (c) retrieval-artifact storage across seven datasets. Each point represents one method. The vertical axis is the unweighted arithmetic mean of the dataset MAP@$k$ values in Table~\ref{tab:main}; each horizontal axis is the geometric mean of the corresponding costs on a logarithmic scale. Parenthesized labels give cost ratios relative to \method{}. Measurement scopes follow Section~\ref{sec:exp:setup}.}
    \label{fig:costs}
\end{figure*}

\subsection{Efficiency and Storage}
\label{sec:exp:efficiency}
Across the seven datasets, \method{} achieves the highest mean MAP ($77.6\%$) and the shortest geometric-mean online retrieval time among the nine methods (Fig.~\ref{fig:costs}). Its retrieval artifacts also take less time to build and occupy less space than those of all six prior TUS methods.

\textbf{Offline.}
\method{} has a geometric-mean build time of $313.4$ seconds, compared with $2{,}179.3$ seconds for \textsf{TACTUS}, a $7.0\times$ reduction (Fig.~\ref{fig:costs}(a)). Its build cost is also lower than those of the other representation- and type-based baselines and \textsf{SANTOS}, whose geometric mean is $2{,}546.7$ seconds. For example, on \textsf{LakeBench\_1k}, \method{} constructs its retrieval artifacts in $1{,}085$ seconds, compared with $66{,}204$ seconds for \textsf{TACTUS}.

\begin{table}[!b]
\centering
\caption{Robustness to candidate-space size on \textsf{LakeBench} (MAP; best per row in \textbf{bold}).}
\label{tab:robustness}
\fontsize{7.5}{9}\selectfont
\renewcommand{\arraystretch}{1.16}
\setlength{\tabcolsep}{1.35pt}
\newcommand{\poolhead}[1]{{\fontsize{7}{8}\selectfont\bfseries #1}}
\begin{tabular*}{\linewidth}{@{\extracolsep{\fill}}llrrrrrrr@{}}
\specialrule{0.8pt}{0pt}{0pt}
\rowcolor{resulthead}
\poolhead{Dataset} & \poolhead{\#Tables} & \poolhead{GenTUS} & \poolhead{TACTUS} & \poolhead{Starmie} & \poolhead{Sherlock} & \poolhead{SATO} & \poolhead{BIRDIE} & \poolhead{Sup-BiEnc.} \\
\specialrule{0.45pt}{0pt}{0pt}
\multirow{4}{*}{LB-1K}
 & 1{,}000  & \cellcolor{bestfill}\textbf{0.627} & 0.426 & 0.270 & 0.394 & 0.405 & 0.457 & \cellcolor{secondfill}0.515 \\
 & 2{,}500  & \cellcolor{bestfill}\textbf{0.634} & 0.432 & 0.291 & 0.411 & 0.412 & 0.451 & \cellcolor{secondfill}0.499 \\
 & 5{,}000  & \cellcolor{bestfill}\textbf{0.627} & 0.432 & 0.303 & 0.391 & 0.385 & 0.437 & \cellcolor{secondfill}0.471 \\
 & 7{,}970  & \cellcolor{bestfill}\textbf{0.602} & 0.409 & 0.304 & 0.368 & 0.364 & 0.430 & \cellcolor{secondfill}0.449 \\
\arrayrulecolor{resultdivider}\specialrule{0.45pt}{0pt}{0pt}\arrayrulecolor{black}
\multirow{4}{*}{LB-30K}
 & 1{,}000  & \cellcolor{bestfill}\textbf{0.651} & 0.417 & 0.272 & 0.415 & 0.422 & \cellcolor{secondfill}0.601 & 0.525 \\
 & 2{,}500  & \cellcolor{bestfill}\textbf{0.637} & 0.429 & 0.282 & 0.398 & 0.415 & \cellcolor{secondfill}0.545 & 0.488 \\
 & 5{,}000  & \cellcolor{bestfill}\textbf{0.632} & 0.427 & 0.299 & 0.388 & 0.399 & \cellcolor{secondfill}0.545 & 0.464 \\
 & 7{,}970  & \cellcolor{bestfill}\textbf{0.613} & 0.417 & 0.297 & 0.364 & 0.373 & \cellcolor{secondfill}0.524 & 0.447 \\
\specialrule{0.8pt}{0pt}{0pt}
\end{tabular*}
\end{table}

\textbf{Online.}
The geometric-mean online time of \method{} is $12.9$ seconds, compared with $77.1$ seconds for \textsf{TACTUS} and $900.5$ seconds for \textsf{SANTOS} (Fig.~\ref{fig:costs}(b)). This gives approximately a $6\times$ reduction relative to \textsf{TACTUS}. \method{} retrieves tables through constrained identifier generation, avoiding a separate candidate lookup and reranking stage. \textsf{BIRDIE} has a geometric-mean online time of $116.3$ seconds, while \method{} achieves a higher mean MAP of $77.6\%$ versus $55.5\%$.

\textbf{Storage.}
\method{} maintains a compact candidate-specific footprint, including its semantic identifiers and retained candidate representations. Its geometric-mean footprint is $7.9$\,MiB, compared with $63.5$\,MiB for \textsf{TACTUS}, $393.6$\,MiB for \textsf{Starmie}, and $3{,}422.1$\,MiB for \textsf{SANTOS} (Fig.~\ref{fig:costs}(c)). The footprint is $8.1\times$ smaller than that of \textsf{TACTUS} and approximately half that of \textsf{BIRDIE} ($15.8$\,MiB), while \method{} achieves higher mean MAP than both methods.

Relative to the lightweight \textsf{Sup-BiEncoder}, \method{} raises mean MAP from $61.3\%$ to $77.6\%$, a gain of $16.3$ percentage points. \method{} achieves this gain with a geometric-mean online retrieval time of $12.9$ seconds, compared with $13.8$ seconds for \textsf{Sup-BiEncoder}. The figure thus places \method{} at the high-quality end of this comparison, with lower costs than all six prior TUS methods.

\subsection{Robustness to Candidate-Space Size}
\label{sec:exp:robustness}

We evaluate retrieval quality over candidate subsets of different sizes from each \textsf{LakeBench} data lake (Table~\ref{tab:robustness}). Models follow the main full-data training setting. At evaluation, we select candidate pools containing $1{,}000$, $2{,}500$, $5{,}000$, or all $7{,}970$ tables from that lake. The full-pool setting matches the \textsf{LakeBench} candidate space in Table~\ref{tab:main}.

\textbf{Robust across pool sizes.}
\method{} achieves the highest MAP at every pool size, ranging from $0.602$ to $0.651$. \textsf{Sup-BiEncoder} is the strongest baseline on \textsf{LB-1K} ($0.449$--$0.515$), while \textsf{BIRDIE} leads the baselines on \textsf{LB-30K} ($0.524$--$0.601$). \method{} maintains stable performance across pool sizes, exceeding the strongest baseline by $5.0$--$15.6$ percentage points.

\subsection{Ablation and Analysis}
\label{sec:exp:ablation}

We first compare four identifier variants to examine how unionability supervision, code structure, and collision handling affect MAP.
We then measure how often unconstrained decoding leaves a result slot without a valid table ID and how many additional relevant tables prefix constraints recover per query.
Finally, we vary beam width to study the trade-off between MAP and query latency.

\begin{table}[!b]
\centering
\caption{Semantic ID ablation. Values are MAP gains of unionability-aware SID over each variant.}
\label{tab:abl-sid}
\fontsize{7.5}{9}\selectfont
\renewcommand{\arraystretch}{1.16}
\setlength{\tabcolsep}{1.6pt}
\begin{tabular}{@{}w{l}{46pt}*{7}{w{r}{\dimexpr(\linewidth-46pt-14\tabcolsep)/7\relax}}@{}}
\specialrule{0.8pt}{0pt}{0pt}
\rowcolor{resulthead}
\textbf{Variant} & \textbf{TUS-S} & \textbf{TUS-L} & \textbf{SAN-S} & \textbf{SAN-L} & \textbf{Wiki} & \textbf{LB-1K} & \textbf{LB-30K} \\
\specialrule{0.45pt}{0pt}{0pt}
\textbf{Ordinary RQ} & +0.266 & +0.031 & $-0.002$ & +0.225\phantom{0} & +0.177 & +0.108 & +0.110 \\
Random      & +0.518 & +0.911 & +0.424  & +0.486\phantom{0} & +0.073 & +0.237 & +0.243 \\
Atomic      & +0.394 & +0.814 & +0.213  & +0.5193 & +0.688 & +0.288 & +0.247 \\
No-Suffix   & +0.118 & +0.167 & +0.058  & +0.238\phantom{0} & +0.246 & +0.299 & +0.342 \\
\specialrule{0.8pt}{0pt}{0pt}
\end{tabular}
\end{table}

\textbf{Semantic identifier.}
We first ablate how the semantic identifier is constructed (Table~\ref{tab:abl-sid}).
We report MAP gains over four variants: ordinary RQ without the unionability objective, random identifiers, atomic identifiers, and No-Suffix.
Atomic identifiers use one unique token per table; No-Suffix drops collision suffixes and selects one table uniformly at random when a base identifier maps to multiple tables.
The improvement over random and atomic identifiers remains substantial on most datasets, reaching $+0.911$ over random identifiers and $+0.814$ over atomic identifiers, confirming that the identifier should encode table semantics rather than be assigned arbitrarily.
The improvement over No-Suffix is consistently positive across all datasets ($+0.058$ to $+0.342$), showing that collision suffixes help keep tables uniquely addressable.
\method{} achieves near-ceiling MAP on both \textsf{TUS\_small} and \textsf{SANTOS\_small}, yet the contribution of unionability supervision differs sharply.
It improves MAP over ordinary RQ by $0.266$ on \textsf{TUS\_small}, while the two variants are within $0.002$ on \textsf{SANTOS\_small}.
The gains over ordinary RQ on \textsf{SANTOS\_large} and \textsf{wiki\_union} are $0.225$ and $0.177$, respectively, and both \textsf{LakeBench} variants gain about $0.11$ MAP.

\begin{figure*}[t]
    \centering
    \includegraphics[width=\textwidth]{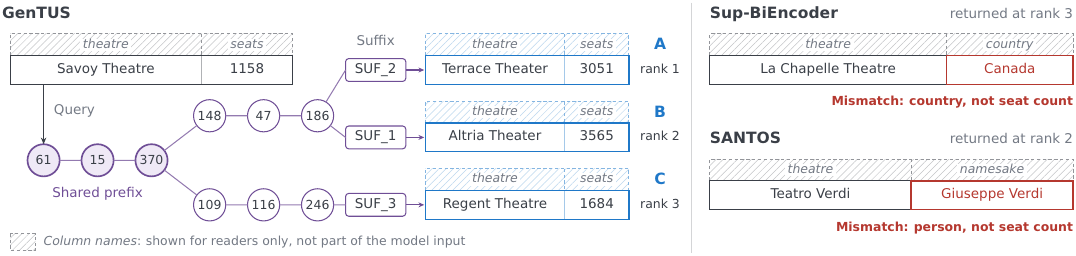}
    \caption{Top-three results returned by GenTUS for a theatre query on \textsf{wiki\_union} with their generated SID paths (left), alongside selected baseline candidates (right). Hatched labels indicate column meanings for illustration.}
    \label{fig:theatre-case}
\end{figure*}

\begin{table}[!b]
\centering
\caption{Prefix-constrained decoding on all datasets. We report the invalid top-$k$ slot rate before constraint and the additional relevant tables recovered per query.}
\label{tab:prefix-decoding}
\fontsize{7.5}{9}\selectfont
\renewcommand{\arraystretch}{1.16}
\setlength{\tabcolsep}{1.6pt}
\begin{tabular}{@{}w{l}{46pt}*{7}{w{r}{\dimexpr(\linewidth-46pt-14\tabcolsep)/7\relax}}@{}}
\specialrule{0.8pt}{0pt}{0pt}
\rowcolor{resulthead}
\textbf{Metric} & \textbf{TUS-S} & \textbf{TUS-L} & \textbf{SAN-S} & \textbf{SAN-L} & \textbf{Wiki} & \textbf{LB-1K} & \textbf{LB-30K} \\
\specialrule{0.45pt}{0pt}{0pt}
Inv. Rate    & 0.005 & 0.004 & 0.128 & 0.604 & 0.242 & 0.379 & 0.395 \\
Added Rel./Q  & 0.000 & 0.033 & 0.580 & 6.257 & 7.480 & 0.533 & 0.588 \\
\arrayrulecolor{resultdivider}\specialrule{0.45pt}{0pt}{0pt}\arrayrulecolor{black}
\textbf{\boldmath$\Delta$MAP}  & 0.000 & 0.001 & 0.057 & 0.403 & 0.198 & 0.021 & 0.022 \\
\textbf{\boldmath$\Delta$Recall} & 0.000 & 0.000 & 0.049 & 0.406 & 0.154 & 0.039 & 0.034 \\
\specialrule{0.8pt}{0pt}{0pt}
\end{tabular}
\end{table}

\textbf{Prefix-constrained decoding.}
We next analyze prefix-constrained decoding (Table~\ref{tab:prefix-decoding}).
The decoder uses a prefix tree over the table identifiers, so each decoding step only expands prefixes that can lead to an existing table.
Unconstrained decoding can spend top-$k$ outputs on invalid identifiers, whereas prefix constraints keep the output list within valid table candidates.
Across all seven datasets, prefix constraints never reduce MAP or Recall.
They have little effect on the two \textsf{TUS} datasets, where unconstrained decoding is already almost always valid.
On \textsf{SANTOS\_large}, $60.4\%$ of the top-20 result slots had no valid table identifier under unconstrained decoding.
With prefix constraints, \method{} retrieved an average of $6.26$ additional relevant tables per query.
MAP increased by $0.403$ on this dataset and by $0.198$ on \textsf{wiki\_union}.
On both \textsf{LakeBench} variants, unconstrained decoding also left many slots without valid IDs, but the constraints recovered fewer than one additional relevant table per query on average.

\begin{samepage}
\textbf{Case study.} The query in Fig.~\ref{fig:theatre-case} pairs theatre names with seat counts, so a unionable table should provide additional theatres with the same information. GenTUS's top three results preserve this pairing. By contrast, Sup-BiEncoder's third result lists theatres and countries, while SANTOS's second result lists theatres and the people they are named after. Both concern theatres, but their second columns cannot extend the query's seat-count column. This illustrates why topical relevance alone is insufficient for table union search.

The SID paths illustrate how GenTUS retrieves multiple unionable tables while keeping them individually addressable. All three results share a three-token prefix. Later tokens separate C, whereas A and B share the same base SID and are distinguished by collision suffixes. These suffixes preserve both tables as separate retrieval targets despite their identical base codes. For this query, GenTUS retrieves 38 labeled positives in its top 40, compared with 12 for Sup-BiEncoder and 10 for SANTOS.\par
\end{samepage}

\textbf{Beam width.}
Finally, we study how beam width trades off retrieval quality against online decoding cost (Fig.~\ref{fig:e12}), using the same checkpoints and query splits as the main effectiveness table.
The quality curve rises sharply at small beams and then quickly saturates: \textsf{SANTOS\_large} is already close to its best MAP by beam $20$, while \textsf{wiki\_union} and the two \textsf{LakeBench} splits reach their plateau around beam $40$--$80$.
Increasing the beam beyond this point brings little accuracy gain on the LakeBench splits, where MAP has already saturated.
Latency generally increases with beam width because the decoder explores more identifier prefixes.
Overall, the figure shows that a moderate beam is sufficient in practice: it recovers almost all of the attainable MAP while avoiding the much higher latency of the largest beams.

\begin{figure}[t]
    \centering
    \includegraphics[width=0.96\linewidth]{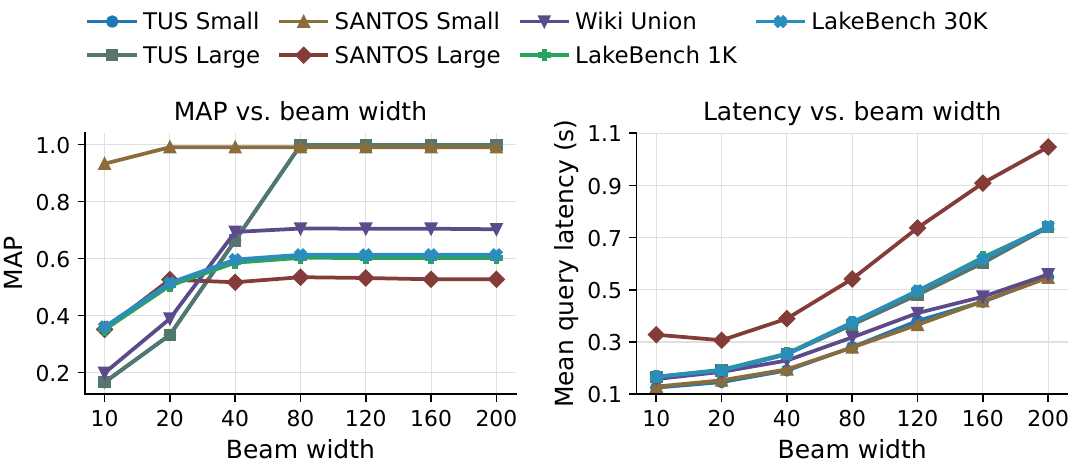}
    \caption{Beam-width trade-off between MAP and mean query latency.}
    \label{fig:e12}
\end{figure}
\endgroup

\section{Conclusion}
\label{sec:conclusion}
We presented \method{}, a generative framework that reformulates table union search as constrained generation over discrete semantic table identifiers.
Offline, \method{} assigns each candidate table a residual-quantized identifier shaped by table-pair unionability annotations, and trains an encoder--decoder generator with a pairwise ranking objective.
Online, a single prefix-constrained generation pass produces the top-$k$ result, with no dense vector index and no per-candidate reranking.
Across seven benchmarks, \method{} achieves the best overall retrieval quality while requiring lower retrieval-artifact construction time, online retrieval time, and retrieval-artifact storage than prior TUS methods.

These results show that generative retrieval is a viable and practical alternative to encode--search--refine pipelines for TUS.
Future work includes extending the formulation to joint unionable-table and joinable-table discovery and evaluating the approach on larger data lakes.

\section*{Acknowledgment}
We used OpenAI Codex for study design, coding, data preparation, analysis, and plots (Sec.~VII) and for copyediting.

\bibliographystyle{IEEEtran}
\bibliography{references}

\end{document}